\documentclass[journal=jctcce,manuscript=article]{achemso}
\setkeys{acs}{usetitle = true}
\usepackage[usenames,dvipsnames]{color}
\usepackage{amsmath}
\usepackage{amsfonts}
\usepackage{graphicx}
\usepackage{mathptmx}
\usepackage{courier} % for typewriter
\normalfont
\newcommand{\onlinecite}[1]{\nocite{#1}\citenum{#1}}

\usepackage[T1]{fontenc}
\usepackage[flushleft]{threeparttable}
\usepackage{multirow}
\usepackage{dcolumn}
\usepackage{bm}
\usepackage{graphicx}
\usepackage[version=3]{mhchem}
\usepackage{braket}
\usepackage{color}
\usepackage{booktabs}
\usepackage{enumerate}
\usepackage{ulem}

\renewcommand{\phi}{\varphi}
\renewcommand{\epsilon}{\varepsilon}

\newcommand{\lin}{\operatorname{span}}
\newcommand{\mat}[1]{\mathbf{#1}}

\renewcommand{\emph}[1]{\textit{#1}}

\newcommand{\markchange}[1]{#1}
\newcommand{\markchangesection}{}

\title{Automated construction of molecular active spaces from atomic valence orbitals}
\date{\today}
\author{Elvira R. Sayfutyarova}
\affiliation{Department of Chemistry, Princeton University, Princeton, NJ 08540, USA}
\affiliation{Division of Chemistry and Chemical Engineering, California Institute of Technology, Pasadena, CA 91125, USA}
\author{Qiming Sun}
\affiliation{Division of Chemistry and Chemical Engineering, California Institute of Technology, Pasadena, CA 91125, USA}
\author{Garnet Kin-Lic Chan}
\affiliation{Division of Chemistry and Chemical Engineering, California Institute of Technology, Pasadena, CA 91125, USA}
\author{Gerald Knizia}
\affiliation{Department of Chemistry, The Pennsylvania State University, University Park, PA 16802,  USA}
\email{knizia@psu.edu}

\begin{document}

\begin{abstract}

We introduce the atomic valence active space (AVAS), a simple and well-defined automated technique for constructing active orbital spaces for use in multi-configuration and 
multi-reference (MR) electronic structure calculations.
Concretely, the technique constructs active molecular orbitals capable of describing \emph{all} relevant electronic configurations emerging from a targeted set of atomic valence orbitals (e.g., the metal $d$ orbitals in a \markchange{coordination complex}).
This is achieved via a linear transformation of the occupied and unoccupied orbital spaces from an easily obtainable single-reference wavefunction (such as from a Hartree-Fock or Kohn-Sham calculations) based on projectors to targeted atomic valence orbitals.
We discuss the \markchange{premises}, theory, and implementation of the idea, and several of its variations are tested.
To \markchange{investigate} the performance and accuracy, we calculate the excitation energies for various transition metal complexes in typical application scenarios.
\markchange{Additionally, we follow the homolytic bond breaking process of a Fenton reaction along its reaction coordinate}.
\markchange{While the described AVAS technique is not an universal solution to the active space problem, its premises are fulfilled in many application scenarios of transition metal chemistry and bond dissociation processes.}
\markchange{In these cases} the technique makes MR calculations easier to execute, easier to reproduce by any user, and simplifies
the determination of the appropriate size of the active space required for accurate results.
% The described technique makes MR calculations easier to execute, easier to reproduce by any user, and simplifies
% the determination of the appropriate size of the active space required for accurate results.
\end{abstract}

\maketitle

\section{Introduction}
Multiconfigurational and multireference (MR) methods remain indispensable in the treatment of 
challenging electronic structure problems. 
Transition metal complexes provide a rich source of examples, as they often feature strongly correlated electronic degrees of freedom, which render density functional theory (DFT) calculations unreliable.
Unfortunately, MR methods require an a priori choice of a suitable set of active molecular orbitals (an {\it active space}), which critically determines the quality of the results.
The choice of active space is non-trivial and often represents a major challenge in practical computations.

The standard way to choose an active space is as follows: First, a full set of molecular orbitals of the molecule is computed with a simple electronic structure method, such as Hartree-Fock (HF) or Kohn-Sham DFT (KS-DFT). Second, these molecular orbitals, both occupied and unoccupied, are  visually inspected, and based on their shape, energy, occupation numbers, etc., one selects into the active space the set of orbitals which are expected to be chemically most relevant
(e.g., with significant transition metal $d$-electron character or ligand character in the case of transition metal complexes---regions which are empirically known to be important).
% But despite the existence of general advice on how to select a good set of starting orbitals \cite{Roos:AS, Roos:AS2,NonDynCorrTMC,QMC}{(\color{red}FIXME: what is ref QMC?)}, there are some problems with this approach.
But despite the existence of general advice on how to select a good set of starting orbitals \cite{Roos:AS, Roos:AS2,NonDynCorrTMC}, there are some problems with this approach.
First, the selection of molecular orbitals for the active space is performed by the user, normally based on personal experience.
%and  subjective criteria.
This makes MR methods hard to apply, and gives results that are complex to reproduce and hard to judge in terms of quality.
This stands in contrast to single-reference calculations which do not require active spaces and therefore do not have this level of arbitrariness. 
Second, the molecular orbitals mix together valence orbitals of different character; 
for example, typically a metal's $d$ atomic valence orbitals contribute to a very large number of molecular orbitals, and it is often not easy to truncate these to a small active subset in such a way that they retain the capability of describing all the right physics in complicated systems.
In the case of large complexes, especially with multiple metal ions, this procedure commonly becomes a matter of trial and error.

Techniques to aid in the construction of high quality active spaces are therefore highly desirable, particularly as more  powerful electronic structure methods are becoming available which are capable of treating increasingly large active spaces efficiently.\cite{sharma:DmrgInPractice,sharma:MpsPt21,sharma:DmrgInQc,sharma:MpsPt22,sharma:IcMrlcc,reiher:DmrgCasScf,reiher:DmrgCdPt2,scuseria:SingledPairedCcOpenShell,scuseria:SeniorityBasedCc,lehtola:PerfectQuadruples,Evangelista:Aci2,Evangelista:Aci1,Evangelista:DSRGPTruncation,Evangelista:MRDSRGPT2,gagliardi:mcpdft1,gagliardi:mcpdft2,thomas2015stochastic,fosso2016large,LiManni1,LiManni2,LiManni3,LiManni4}
There have been a number of contributions in this area, and the general procedures can loosely be summarized as based on
 estimating the (correlated) occupation numbers of the orbitals (including the full single orbital density matrix in Ref.~\onlinecite{Keller:ActiveSpaceSelection,Stein:ActiveSpaceSelection}), followed by selecting the active orbitals based on partial occupancy according to an input threshold. The various techniques differ primarily 
in how the occupation number information is obtained. For example, correlated occupation numbers
can be estimated from unrestricted Hartree-Fock~\cite{pulay1988uhf,bofill1989unrestricted} or Kohn-Sham calculations or from correlated calculations,
such as MP2~\cite{jensen1988second,abrams2004natural} or approximate DMRG calculations~\cite{wouters2014communication,Stein:ActiveSpaceSelection,Stein2,stein:ActiveSpaceSelectionForReactions}.

\markchange{Nonetheless, while these existing automated approaches are advancements from typical ad hoc active space constructions}, there is still room for further improvement.
For example, an obvious drawback of the above procedures is that they all require a non-trivial preliminary calculation:
either a correlated calculation must be performed, or a suitable broken symmetry solution must be found.
This is not always possible: for example, there may not always be a broken symmetry solution involving the region of interest, or
the preliminary correlated calculation may simply be too costly. 
A perhaps less obvious drawback, but one which one encounters in practice, is that there is no guarantee that the active orbitals
found in these automated procedures are actually spatially located in the region of chemical interest. For example, in 
models of enzymatic binding sites, particularly those with charged ligands, the unpaired electrons may lie in functional groups which are spatially 
far from and irrelevant to the chemistry of the metal center. In this case, an additional inspection is once again required to choose the subset of
active orbitals of chemical interest.

Here we propose an alternative approach to construct molecular active spaces for multireference problems automatically and systematically.
This approach does not suffer from the drawbacks mentioned above, and is particularly well-suited for practical calculations involving 
transition metal complexes. In its simplest formulation, the procedure  requires only an easily obtained single-determinant reference function, together with a choice of target atomic valence orbitals.
The technique is based on the following idea: It is empirically known that one can typically identify a small set of atomic valence orbitals which give rise to the strong correlation effects (for example, $d$ atomic valence orbitals in transition metal complexes). 
We therefore aim to construct a set of active molecular orbitals by defining them in terms of these atomic valence orbitals.
Concretely, using simple linear algebra, we can define mathematical rotations of the occupied and virtual molecular orbitals \markchange{of the single-determinant reference function} which maximizes their given atomic valence character (e.g. $3d$ character). 
After this rotation, the relevant molecular orbitals to include into the active space can be selected automatically. 
\markchange{The ideas employed here are closely related to earlier work of Iwata\cite{iwata:ValenceTypeVacantOrbitalsForCiCpl1981} and Schmidt and coworkers\cite{schmidt:ValenceVirtualOrbitalsJPCA2015}  regarding the construction of approximate valence spaces;
however these earlier works focused on constructing molecular orbitals capable of spanning the \emph{entire} valence space (not only the valence space of selected atomic orbitals, with the goal of identifying active spaces for processes involving said atomic orbitals), and the latter work also markedly differs in intent.
Similar mathematical techniques were also used in the construction of molecule-intrinsic minimal basis sets capable of spanning a set of given occupied orbital spaces\cite{ruedenberg:FullValenceCasLocalization1,ruedenberg:FullValenceCasLocalization2,lee:PolarizedAtomicOrbitals1,HeadGordon:PolarizedAtomicOrbitals2,auer:EnvelopingLocalizedOrbitals,Ruedenberg:MoleculeIntrinsicMinimalBasis,subotnik:LocalizedVirtuals,ruedenberg:ExactRepresentationOfRdm,glezakou:so2,laikov:IntrinsicBasis,knizia:iao,west:quambo2013}.}

Sec.~\ref{sec:Theory} explains the motivation, background, theory, and implementation of the construction proposed here.
Sec.~\ref{sec:Results} then describes criteria for judging the quality of the constructed active spaces, and discusses its application to a large number of prototypical MR calculations of metal complexes.
Sec.~\ref{sec:Conclusions} describes conclusions and possible implication for future work.

\section{Theory}\label{sec:Theory}

\subsection{What is an active space?}\label{sec:WhatIsAnActiveSpace}

Many naturally occuring stable molecules have an electronic structure which is qualitatively well described by a single-determinant self-consistent field (SCF) wave function, 
such as used in Kohn-Sham Density Functional Theory (KS-DFT) or Hartree-Fock (HF).
In this case, the entire space of molecular orbitals (MO) is strictly divided into fully occupied and 
fully unoccupied molecular (spin-) orbitals.
However, there are important classes of chemical systems in which this picture breaks down and where a superposition of multiple $N$-electron determinants is required to describe the electronic structure even qualitatively.
This phenomenon is sometimes called \emph{strong correlation}; prototypical cases in which it occurs are (a)  the process of homolytic bond breaking, and (b) various kinds of transition metal complexes---particularly when the complex is in an overall low-spin state generated
by coupling of the metal to other metals or redox non-innocent ligands, as frequently encountered in catalysis and in bio-inorganic systems.

Both of these prototypical cases share the same root cause: The occurrence of valence atomic orbitals with energy levels similar to other valence orbitals, but with poor orbital overlap, 
giving rise to small energy splittings between bonding and anti-bonding linear combinations.
In these cases quantum resonances between both bonding and anti-bonding orbitals must be explicitly considered to qualitatively describe the electronic structure---and single-determinant wave functions are incapable of doing so, because in these each MO either \emph{is} occupied or \emph{is not}, but not both.
The core idea developed in this manuscript is that in these two most important cases, the emergence of strong correlation is tightly linked to \emph{specific valence atomic orbitals, which are easy to identify}. In the case of transition metals, these are the compact $d$ orbitals (and possibly some specific ligand orbitals), and in the case of bond dissociation, these are the valence atomic orbitals of the dissociating atom(s).
Thus, in the following we will assume that a small number of specific valence atomic orbitals are explicitly selected by the user (e.g., the $d$ orbitals of metal centers), and that the goal is to construct an active space suitable for describing all highly relevant determinants that they give rise to.
We stress that this initial selection is easy in practice---the core problem is how to use the information to build a suitable active space.

We thus briefly consider what an active space \emph{is}. 
An active space wavefunction is one where the superpositions
of determinants are restricted so that varied occupations are found only within the active orbitals $\{ \phi_1 \ldots \phi_m\}$. Technically, this means that the wavefunction can be written as 
a second quantized product
\begin{align}
|\Psi\rangle = \{ \phi_1 \ldots \phi_m \} | \mathrm{core} \rangle \label{eq:actwf}
\end{align}
where $\{ \phi_1 \ldots \phi_m\}$ denotes a general occupancy wavefunction within the active orbitals, and $|\mathrm{core}\rangle$
denotes a single determinant. 
It is clear that the active orbitals must span the space of our chosen specific valence atomic orbitals. However, 
Eq.~(\ref{eq:actwf}) additionally implies that the rest of the molecule must be well described by the single core determinant.
To achieve this, the active space must contain orbitals additional to our set of specific valence atomic orbitals, and
to remain compact, we need to define the \emph{minimal} additional set.

To this end, we here employ the basic observation in density matrix embedding theory (DMET),\cite{GK_GKC1,GK_GKC2,zheng2016ground,Wouters2016,scuseria,scuseria2,troy,troy2} which 
describes how to construct such an active space explicitly.
In particular, DMET tells us that the active space with the above properties is at most twice the size of the initial set of chosen valence atomic orbitals. 
However, in this work we  modify the presentation and practice of the DMET procedure
to make it more natural in the active space setting.
In particular, in contrast to the original presentation in references\onlinecite{GK_GKC1,GK_GKC2} (but as described in the appendix of Ref.~\onlinecite{zheng2016ground})
we do not introduce separate ``fragment'' and ``bath'' orbitals, but rather retain the occupied and virtual character of the constructed active orbitals.
This has the important benefit that it leads to a natural truncation procedure, which allows us to further reduce the size of the active space in
the most chemically meaningful way.
In Sec.~\ref{sec:EntangledOrbsAsActiveSpace} we describe the isolation of entangled orbitals for the active space construction, and Sec.~\ref{sec:TechnicalDetails} will provide technical details and discuss various practical aspects relevant in the active space case.
\markchange{Finally, Sec.~\ref{sec:BeyondAvas} outlines how the presented approach may be extended to more complex cases, such as double-shell effects or complex metal/ligand interactions.}

\subsection{Isolating target-overlapping orbitals for the active space}\label{sec:EntangledOrbsAsActiveSpace}

Let $A=\{\ket{p}\}$ denote the (small) set of chosen target valence
atomic orbitals (not necessarily orthonormal), which we expect to be
responsible for strong correlations (e.g. the five $3d$ orbitals of a
third row transition metal atom of a complex, details on their
selection and representation will follow).  Let $|\Phi\rangle$ denote
a Slater determinant, which represents the electronic structure of our
system of interest at the SCF level (HF or KS-DFT).

Our method for active space construction is built around the following
physical assumptions, which are those used in 
DMET\cite{GK_GKC1,GK_GKC2,qmmmdmet,Wouters2016}:(a) a
SCF wave function $\ket{\Phi}$ may be unable to describe
precisely how our target atomic orbitals are bonded with the rest of
the molecule; however, it will generally describe the \emph{rest} of
the molecule reasonably \markchange{well} (experience in DMET suggest that this works
even in cases where the SCF wave function as a whole is qualitatively
\emph{very} wrong\cite{GK_GKC2}), and (b) we can isolate the part of
$\ket{\Phi}$ which involves our target AOs from the part which does
not, and employ the former part as the active space (therefore allowing it
to be replaced by a more powerful wave function description), and retain the
simple determinantal description of rest.

To this end, we employ a rotation within the set of occupied molecular
orbitals of $\ket{\Phi}$ which splits them into two groups: one group
which has overlap with our target AOs, and one group which does not.
A simple dimensional counting argument will show that for a set of
$|A|$ selected target AOs, there is a rotation of the occupied
molecular orbitals such that at most $|A|$ of them have non-zero
weight on the target AOs.  We similarly split  the set of virtual
molecular orbitals into one group of at most $|A|$ virtual orbitals
which have weight on the target atomic valence orbitals, and the other
group which does not.  The idea is now to explicitly construct these
rotated orbital groups, and then employ the at most $2|A|$ combined
occupied and virtual orbitals with target overlap as active orbitals,
while leaving the other occupied and virtual orbitals without target
overlap as inactive (closed-shell) or virtual orbitals, respectively,
in the following multi-configuration treatment.  As all of the
selected AOs in $A$ then lie within the span of this active space, the
resulting multi-configurational wave function is then capable of
representing arbitrary quantum resonances involving the target AOs.

We first discuss the occupied case.  Let $i = 1\ldots N_\mathrm{occ}$
denote $N_\mathrm{occ}$ occupied molecular orbitals (MOs) of
$|\Phi\rangle$.  The projector $\hat P$ onto the space of atomic
orbitals in $A$, is given by
\begin{align}
  \hat P &= \sum _{p,q\in A} |p \rangle [\sigma^{-1}]_{pq} \langle q |.
\end{align}
Here $\sigma$ denotes the $|A|\times|A|$ target AO overlap matrix with
elements $[\sigma]_{pq} = \langle p|q\rangle$, and $\sigma^{-1}$ its
matrix inverse.  \

Employing these projectors, we construct the first set of active
orbitals by rotating $\ket{\Phi}$'s occupied MOs $\{\ket{i}\}$ as
follows.
First, we calculate the $N_\mathrm{occ}\times N_\mathrm{occ}$ overlap
matrix of occupied orbitals projected onto $\lin(A)$, the space of
selected target atomic orbitals:
\begin{align}
 [\mat S^{A}]_{ij} = \langle i|\hat P|j \rangle,
 \label{eq:SA}
\end{align}
where $i,j$ are occupied orbital indices.  Next we compute the
$N_\mathrm{occ}\times N_\mathrm{occ}$ (unitary) matrix of eigenvectors
$[\mat U]_{ij}$ of $\mat S^{A}$, such that
\begin{align}
   \mat S^{A} \mat U = \mat U
   \operatorname{diag}(\sigma_1,\ldots,\sigma_{N_\mathrm{occ}})
\end{align}
(where $\operatorname{diag}(\ldots)$ denotes a diagonal matrix of the
given elements), or, written in component form,
\begin{align}
   \forall i,j:\;\sum_k [\mat S^A]_{ik} [\mat U]_{kj} = [\mat U]_{ij}
   \sigma_j.
\end{align}
There are at most $|A|$ non-zero eigenvalues $\{\sigma_i\}$, because
$\lin(A)$ is a $|A|$-dimensional space, and $[\mat S]^{A}_{ij}$
involves a projection onto it.  Furthermore, the eigenvectors $[\mat
  U]_{ij}$ of $S^{A}_{ij}$ define a rotation on the occupied orbitals:
\begin{align}
   \ket{i} \quad\mapsto\quad \ket{\tilde i} = \sum_{k} \ket{k} [\mat
     U]_{ki}, \label{eq:RotatedOccupiedOrbs}
\end{align}
which clearly separates them into two groups: The at most $|A|$
rotated occupied orbitals $\ket{\tilde i}$ with $\sigma_i\neq 0$,
which have non-vanishing overlap with our target atomic orbitals (and
which therefore should go into the active space), and
the remaining $\ket{\tilde i}$ with $\sigma_i= 0$ which have \emph{no}
overlap with our target atomic orbitals, and therefore can stay as
inactive (inner closed shell) orbitals in the subsequent
multiconfigurational methods.

Note that the rotated occupied orbitals $\{\ket{\tilde i}\}$ in
\eqref{eq:RotatedOccupiedOrbs} are obtained as a unitary
transformation of $\ket{\Phi}$'s original occupied orbitals
$\{\ket{i}\}$.  Consequently, a determinantal wave function
$\ket{\tilde\Phi}$ built from the $\{\ket{\tilde i}\}$ is physically
equivalent (differs by at most a phase factor) from the original
determinant $\ket{\Phi}$.  That is, so far we have done nothing to
$\ket{\Phi}$ except for splitting its occupied orbitals into a convenient
set of at most $|A|$ orbitals related to our $|A|$ target AOs and the
remaining set we can treat as inactive.

We then proceed similarly for the virtual orbitals $\{\ket{a}, 1 =
1\ldots,N_\mathrm{vir}\}$ of $\ket{\Phi}$: We form the
$N_\mathrm{vir}\times N_\mathrm{vir}$ projected overlap matrix
\begin{align}
 [\bar {\mat S}]^{A}_{ab} = \langle a|\hat P|b \rangle,
 \label{eq:SAvirt}
\end{align}
where $a,b$ are virtual orbital indices, then find its unitary matrix
of eigenvectors $\bar{\mat U}$ such that
\begin{align}
   {\bar {\mat S}}^{A} \bar{\mat U} = \bar{\mat U}
   \operatorname{diag}(\sigma_1,\ldots,\sigma_{N_\mathrm{vir}})
   \qquad\Leftrightarrow\qquad \forall a,b:\;\sum_c [{\bar{\mat
         S}}^A]_{ac} [\bar{\mat U}]_{cb} = [\bar{\mat U}]_{ab}
   \sigma_b,
\end{align}
and use $\bar{\mat U}$ to rotate the virtual orbitals via
\begin{align}
   \ket{a} \quad\mapsto\quad \ket{\tilde a} = \sum_{c} \ket{c}
       [\bar{\mat U}]_{ca}. \label{eq:RotatedVirtualOrbs}
\end{align}
Again, the at most $|A|$ of the new virtual orbitals $\{\ket{\tilde
  a}\}$ with eigenvalues $\sigma_a\neq 0$ are selected for the active
space, while the remaining orbitals will stay unoccupied in the
subsequent multi-configuration treatment.

Finally, having active orbitals with overlap with $A$ from both sides,
occupied and unoccupied orbitals in $|\Phi\rangle$, we can form the
total active space by combining the sets of $\{\ket{\tilde i}\}$ and
$\{\ket{\tilde a}\}$ with non-zero projected overlap eigenvalues.
Since the combined set includes \emph{all} orbitals which have
non-vanishing overlap with our target space $\lin(A)$, all of the
selected AOs in $A$ then lie within the span of this active space.
Therefore, a multi-configurational wave function with this active
space will be capable of representing arbitrary quantum resonances
involving the target AOs\markchange{---}which was the goal of our construction.

\subsection{Technical details of the construction}\label{sec:TechnicalDetails}
Sec.~\ref{sec:EntangledOrbsAsActiveSpace} discusses the formal framework of the active space construction.
However, several practical aspects still need to be discussed:
(a) How are the target AOs $A=\{\ket{p}\}$ chosen and represented?
(b) How are the actual rotation matrices $\mat U$ (eq.~\eqref{eq:RotatedOccupiedOrbs}) and $\bar{\mat U}$ (eq.~\eqref{eq:RotatedVirtualOrbs}) computed in practice?
(c) Can the active space (formally twice the number of the target AOs) be further reduced in size?
(d) How should open-shell systems be handled? (In particular, what to do for restricted open-shell functions $\ket{\Phi}$?)
We will discuss these questions in the current and next subsections.

Let us first assume that $|\Phi\rangle$ is a closed-shell Slater determinant obtained from an SCF calculation. 
Its occupied and virtual molecular orbitals are expressed as:
\begin{align}
   |i \rangle &= \sum_{\mu \in B_1}  |\mu \rangle C^{\mu}_i\notag
\\ |a \rangle &= \sum_{\mu \in B_1}  |\mu \rangle \bar C^{\mu}_a, \label{eq:MOexpan}
\end{align}
where $\mu$ are basis functions from the (large) computational basis set $B_1$ (e.g., cc-pVTZ or def2-TZVPP), and $C^{\mu}_i=[{\mat C}_\mathrm{occ}]_{\mu i}$ and $\bar C^{\mu}_a=[{{\mat C}}_\mathrm{vir}]_{\mu a}$ are the coefficients of the basis function $\mu$ in the expansion of the occupied orbital $i$ and virtual orbital $a$, respectively.
${\mat C}_\mathrm{occ}$ and ${{\mat C}}_\mathrm{vir}$ denote the $|B_1|\times N_\mathrm{occ}$ occupied and $|B_1|\times N_\mathrm{vir}$ virtual sub-matrices of the $|B_1|\times |B_1|$ SCF orbital matrix $\mat C$ (note that $N_\mathrm{occ}+N_\mathrm{vir}=|B_1|$---each orbital is either occupied or virtual).

In general, computational basis sets such as $B_1$ do not contain basis functions directly corresponding to AOs of any sort.
For this reason, we here select our target AOs $A=\{\ket{p}\}$ based on a second auxiliary basis set $B_2$. 
This is a minimal basis set of tabulated free-atom AOs (here MINAO is used\cite{knizia:iao}; \markchange{but other choices, such as subsets of ANO-RCC,\cite{widmark:AnoRccHHe,roos:AnoRccMainGroups} ano-pVnZ,\cite{neese:ano2011} or ANO-VT-XZ\cite{ANO_VT}, could be considered; see also Sec.~\ref{sec:BeyondTheValenceSpace}}).
This choice leads to  simple expressions for the projected overlap matrices
\begin{gather}
S^A_{ij} = \braket{i|\hat P|j} = \sum_{\mu \mu'}  C^{\mu}_i P_{\mu \mu'} C^{\mu'}_j \label{eq:SijProjectorsDirect} \\
{\bar S}^A_{ab} = \braket{a|\hat P|b} = \sum_{\mu \mu'}  {\bar C}^{\mu}_a P_{\mu \mu'} {\bar C}^{\mu'}_b, \label{eq:SabProjectorsDirect}
\end{gather}
where the matrix elements of the projector are
\begin{equation}
  P_{\mu \mu'} =\sum_{pp'} \langle \mu|p \rangle [\sigma^{-1}]_{pp'}  \langle p'| \mu' \rangle.
\end{equation}
Combining all formulas into a numerical algorithm, the rotated orbitals are constructed as follows:
\begin{itemize}
 \item Let $ A \subset B_2$ denote the subset of AOs we choose as target AOs for the active space construction (for example, the five $3d$ AOs in a transition metal complex with one metal center). 
 \item Form the overlap matrix $\mathbf{\sigma}$ with elements $\sigma_{pp'} = \langle p | p' \rangle$, where $p, p' \in A$\markchange{, as well as
 $\mathbf{\sigma}$'s} inverse matrix, with elements $\sigma^{pp'}=[\sigma^{-1}]_{pp'}$. Both matrices have dimension of $|A| \times |A|$.
 \item Form the overlap matrix $\mathbf{S}_{21}$ between the functions $p$ of $A \subset B_2$ and the functions $\mu$ of the large basis set $B_1$, with elements $[\mat S_{21}]_{p\mu}= \langle p | \mu \rangle$.
 \item Form the projector $P_{\mu \mu'} =\sum_{pp'} \langle \mu|p \rangle \sigma^{pp'}  \langle p'| \mu' \rangle$ , or $\mathbf{P}=\mathbf{S}_{21}^\dagger \mathbf{\sigma}^{-1}\mathbf{S}_{21}$.
 \item Form the projected overlap matrices $\mathbf{S}^A={\mat C}_\mathrm{occ}^\dagger\mat P\,{\mat C}_\mathrm{occ}$ for the occupied orbitals (eq.\;\eqref{eq:SijProjectorsDirect}), and $\bar {\mat S}^A={\mat C}_\mathrm{vir}^\dagger \mat P\, {\mat C}_\mathrm{vir}$ for the virtual orbitals  (eq.\;\eqref{eq:SabProjectorsDirect}).
\item Finally, diagonalize both projected overlap matrices to obtain the transformation matrices separating the MO sets by overlap with $\lin(A)$. 
Concretely, diagonalize $\mathbf{S}^A$ to obtain the eigenvector matrix $\mat U$, and use it to find the transformed occupied orbital matrix ${\tilde{\mat C}}_\mathrm{occ} = {{\mat C}}_\mathrm{occ} \mat U$. 
Then diagonalize $\bar{\mathbf{S}}^A$ to obtain the eigenvector matrix $\bar{\mat U}$, and use it to find the transformed virtual orbital matrix ${\tilde{\mat C}}_\mathrm{vir} = {{\mat C}}_\mathrm{vir} \bar{\mat U}$. These are the expansion coefficients of $\{\ket{\tilde i}\}$ (eq.\;\eqref{eq:RotatedOccupiedOrbs}) and $\{\ket{\tilde a}\}$ (eq.\;\eqref{eq:RotatedVirtualOrbs}), respectively.
\end{itemize}

Rather than using the minimal basis $B_2$ directly, one could consider choosing the target AOs from a set of polarized AOs which take the molecular environment into account, such as the Intrinsic Atomic Orbitals (IAOs).\cite{knizia:iao}
If the IAOs are given as
\begin{align}
  |\tilde{p}\rangle = \sum_{\mu} |\mu\rangle T_{\mu p},
\end{align}
where $T_{\mu p}$ denotes the elements of the $|B_1|\times |B_2|$ IAO transformation matrix,\cite{knizia:iao} this can be incorporated by updating the projection matrix in $S^A_{ij}$ and $S^A_{ab}$ in eqs.~\eqref{eq:SijProjectorsDirect} and \eqref{eq:SabProjectorsDirect} as
\begin{align}
  \mathbf{P} = \mathbf{S}\mathbf{T}(\mathbf{T}^{\dagger} \mathbf{S} \mathbf{T})^{-1}\mathbf{T}^\dagger\mathbf{S}.
\end{align}
For simplicity, we do not follow this approach in this work, and choose the target AOs directly from the minimal basis $B_2$ as described above. 

\markchange{Additionally, in some cases one may wish to extend the target space by some AOs which lie outside the valence space. This scenario is discussed in Sec.~\ref{sec:BeyondTheValenceSpace}.}

% \markchange{Note: If one wants to include additional AOs, which are not usually considered as valence AOs, but are believed 
% to be important for correlation, one should use ANO-RCC, ANO-VT-XZ or another non-minimal ANO basis set.
% For example, the ``double-shell' effect can be easily included in the AVAS method by including both 3$d$ and 4$d$ AOs in the ``target'' space of the active space construction
% using ANO-RCC basis set.}

\subsection{Truncating the active space}

The eigenvalues $\sigma_i$ and $\sigma_a$ of the projected overlap matrices 
in eqs. \eqref{eq:RotatedOccupiedOrbs} and \eqref{eq:RotatedVirtualOrbs} 
reflect the degree to which the transformed orbitals $\ket{\tilde i}$  and $\ket{\tilde a}$ overlap with the space of our target AOs.
If we include every such transformed orbital with $\sigma_a \neq 0$ and
$\sigma_i \neq 0$ into our active space, 
%even if the eigenvalue is non-zero, but very small, then 
the resulting CAS space will exactly include all electronic configurations which can be formed over the given AOs and the maximum size of the CAS space is twice
that of the target AO space. However, this CAS space may be too large and we may need to truncate it.
Here we can use the fact that often many of the $\sigma_i$ and $\sigma_a$ are small.
As a practical measure, we can set a threshold, such as 0.05--0.1, to exclude MOs with negligible overlap with $\lin(A)$.
In addition to reducing the size of the active space, this can further 
improve the reproducibility of calculations in the case of very small eigenvalues. 
This threshold becomes the only numerical parameter to be chosen by the user\markchange{, and together with the selection of target AOs and the type of the SCF wave function $\ket{\Phi}$, it fully determines the active space}.

Of course, if truncation is used, the active space no longer captures all possible configurations involving the target AOs which can be formed. However, the truncation does {\it not} affect the quality
of the description of the rest of the molecule by the core determinant in Eq.~(\ref{eq:actwf}). This guarantees
that the CASCI energy lies below the variational HF energy. We can also imagine the opposite tradeoff,
where one obtains a truncated active space which retains the ability to capture all possible configurations involving the target AOs, at the cost of worsening
the quality of the core determinant which describes the rest of the molecule.
(In DMET language, this would correspond to truncating the ``bath'' orbitals,
which is considered in Refs.~\onlinecite{Wouters2016} and \onlinecite{qmmmdmet}).
However, this may be a worse truncation procedure in the current setting, as the energy gained by treating the fluctuations in occupation number in the target AOs (such as a TM
$3d$ shell) may not make up for the energy lost in incompletely describing the mean-field hybridization between the target AOs and the rest of the molecule.  
In particular, this second bath truncation procedure can, in principle, lead to a CASCI energy above the variational HF energy.

\subsection{Treatment of open-shell systems}
\label{sec:openshell}

If $\ket{\Phi}$ is a closed-shell determinant, then the active space construction algorithm can be directly used as described in Sec.~\ref{sec:TechnicalDetails}.
However, in the case of open shell determinants, several choices can be considered:
\begin{enumerate}
  \item One may perform the algorithm separately for alpha and beta orbitals, thus creating active orbitals with different spatial parts for alpha and beta-spin electrons.
While this choice is the most straight-forward and, arguably, creates the best initial active orbitals in the open-shell case, this option is not directly feasible if a spin-adapted multiconfigurational calculation will follow---most existing MCSCF programs cannot use such unrestricted orbitals (although this is implemented in the code we use here~\cite{pyscf}).
A possible remedy for this problem would be to construct a single set of ``corresponding orbitals''\cite{amos1961single} from the separate alpha- and beta-orbital sets, but this has not been tested here.
\item \label{onlyalpha} One may use \emph{exclusively} the alpha
orbitals to construct the active space orbitals (and inactive orbitals determining the core determinant).
This treatment can be applied to both restricted and unrestricted SCF functions $\ket{\Phi}$ in a simple manner.
The rationale for this is that in the restricted open-shell case, the occupied beta orbitals lie entirely within the linear span of the occupied alpha orbitals, so one can argue that this choice takes care of both spin cases.
However, this argument is somewhat misleading because it may lead to some \emph{unoccupied} beta orbitals \markchange{being} transformed into the core space, therefore enforcing their occupation with two electrons.
This error in the core means that the CASCI energy may be higher than the variational HF energy.
If the singly-occupied MOs in $\ket{\Phi}$ have only small components on the target valence AOs, this can lead to very bad CASCI wavefunctions.

\item \label{openshell} If a ROHF determinant $\ket{\Phi}$ is used, one may apply the construction of Sec.~\ref{sec:TechnicalDetails} exclusively to the doubly-occupied and fully unoccupied orbitals of $\ket{\Phi}$, to form the core determinant and initial part of the active space, and then include \emph{additionally} all the singly-occupied orbitals of $\ket{\Phi}$ into the active space.
The CASCI energy is then guaranteed to be below the variational HF energy, and further spin-adaptation can be used.
The main drawback is that the active space is usually larger in this procedure.

\end{enumerate}
By default, we use method \ref{onlyalpha}, however, we compare the different schemes in one of the systems below.

\section{Computational Details}\label{sec:ComputationalDetails}
\label{sec:results}

We implemented the atomic valence active space (AVAS) construction within the \textsc{PySCF}\cite{pyscf} package. 
All Complete Active Space Self-Consistent Field (CASSCF), Complete Active Space Configuration Interaction (CASCI), and strongly contracted $N$-electron valence state perturbation theory (NEVPT2)\cite{nev1,nev2,nev3,Sheng2016}
calculations were carried out using \textsc{PySCF}. 
For active spaces with more than 16--17 orbitals we used the \textsc{Block} code\cite{Sharma2012} through the  \textsc{PySCF-Block} interface to perform DMRG calculations in the active space.
The AVAS construction is also being implemented in a development version of Molpro.

For simplicity, we used all-electron cc-pVTZ-DK\cite{Dunning:ccpVnZ_HBCOFNe,Peterson:cc_pVDZ} basis sets for all systems, apart from the Fenton reaction (\emph{vide infra}).
For the auxiliary minimal basis $B_2$ used to choose the target AOs, we employed the MINAO basis\cite{knizia:iao}, which is a truncated subset of the cc-pVTZ basis; for most atoms, this set 
consists of spherically averaged ground-state Hartree-Fock orbitals for the free atoms.
We did not use point group symmetry in the present calculations, since in a straight-forward implementation, the MOs do not necessarily retain symmetry-adaption after rotation; however, if symmetry-adaptable sets of target AOs are chosen, symmetry respecting orbital rotations can in principle be constructed.
%% \cgk{regarding the symmetry adaption: I made an implementation of this for Molpro (not checked in) which retains the symmetry adaption. I do not think there is a fundamental issue with this. At least not with D2h subsets.}
For completeness, scalar relativistic effects were included using the exact-two-component (X2C) approach,\cite{Liu2006,x2c} but this did not lead to significant differences from
the non-relativistic calculations in any of the considered examples. Spin-orbit coupling was not considered.

\markchange{
For the case of the Fenton reaction, the geometries were optimized with the DSCF and GRAD modules of Turbomole 7.0. This was done at the level of symmetry-broken unrestricted B3LYP\cite{becke:b3lyp} with def2-TZVP basis sets\cite{Weigend:def2SVP_def2TZVPP}, starting from the structures provided in Ref.~\onlinecite{petit:FentonReaction}. The RI approximation was not employed, and solvation effects were not considered.
The characters of the starting geometry, transition state geometry, and product geometry were confirmed by computing the analytic nuclear Hessians at these points, via the AOFORCE module.\cite{deglmann:aoforce}
The reaction path was computed by tracing the intrinsic reaction coordinate\cite{fukui:irc} in both directions,
% of the lowest eigenvector of the mass weighted Hessian,
starting at the transition state geometry.
This was done using Turbomole's DRC module.
Finally, the structures of the starting point, transition state, product, and IRC segments (of both directions) were joined, aligned, and compressed using the development version of IboView.\cite{knizia:IboViewHp,knizia:CurlyArrows}
% Finally, the structures of the starting point, transition state, product, and IRC segments (of both directions) were joined and aligned using the development version of IboView, and geometry frames closer than 0.5 bohr*amu^{1/2} were deleted in the IRC segments to reduce the number of final geometry frames.
Employing these geometries, the reported multi-configuration AVAS calculations were performed with \textsc{PySCF}, using  cc-pVTZ orbital basis sets and a non-relativistic Hamiltonian.}

The used molecular geometries are supplied in the supporting information, as are selected visualizations of the obtained active spaces.
Orbital visualizations were made with IboView,\cite{knizia:IboViewHp,knizia:CurlyArrows} and show iso-surfaces enclosing 80\% of the orbital's electron density.

\section{Results and Discussion}\label{sec:Results}
% Judging the quality of an active space is not trivial.
Judging the quality of an initial active space is \markchange{difficult}.
Here we employ two complementary criteria:
\begin{enumerate}
% \item In the course of a CASSCF calculation, the active space orbitals are optimized.
% If the overlap of the optimized active space with our initial active space guess remains high, \markchange{it means that the initial active space is already capable of representing the energetically most important electronic configurations of the system---so} our active space guess is of high quality.
\item In the course of a CASSCF calculation, the active space orbitals are optimized.
If the overlap of the optimized active space with our initial active space guess remains high, \markchange{we here regard} our active space guess \markchange{as being} of high quality. To quantify this aspect, we compute the $N_\mathrm{act}\times N_\mathrm{act}$ overlap matrix
\begin{align}
   \mat S_\textrm{change} = ({\mat C}_\mathrm{act}^\mathrm{final})^\dagger \mat S ({\mat C}_\mathrm{act}^\mathrm{initial})
\end{align}
between the initial guess and the optimized final active orbitals, and compute its singular value decomposition.
If all singular values are close to 1.0, the active space remains mostly unchanged \markchange{during the optimization}. 
On the contrary, each singular value close to 0.0 indicates that an initial active orbital
had to be completely replaced by an unrelated orbital. \markchange{The latter case may indicate that the initial active space lacks the capability to represent some essential features of the strongly correlated electronic structure of the given molecule (since the optimized active orbital, which presumably is needed to represent energetically important electron configurations of the system, lies completely outside the linear span of the initial active space guess); the initial active space guess may therefore need to be changed or enlarged.}
%  \markchange{(lying outside of the span of the initial active space)}
\item We compare our computed excitation energies to experimental results and other high-level calculations reported in the literature.
When constructing a well-behaved series of active spaces, we should be able to see convergence or stability of the computed properties
with respect to the active space size.
\end{enumerate}
\markchange{We note that, strictly speaking, neither criterion can establish an initial active space as \emph{``definitely good''}: While criterion 1 tests whether the CASSCF-optimized active space is close to the initial active space, there is no formal guarantee that the optimized CASSCF wave function itself is the best representative of the sought after electronic state (e.g., the optimization algorithm could be stuck in an undesirable local minimum). While criterion 2 establishes the compatibility with some specific physical properties and reference systems, it cannot guarantee that no other physical properties or systems exist which need different active spaces for a proper description.
% However, both criteria \emph{could} establish the premise of our active space construction \emph{to be violated}---by showing counter-examples and negative results.
However, both criteria \emph{could} establish our active space construction \emph{to be unfit} for its target applications---by showing counter-examples and negative results.
% Ultimately, the construction will be judged by its usefulness in practical applications.
% Ultimately, the construction will have to be judged by its usefulness in practical applications.
% Unfortunately, the true versatility and range of applicability of an active space construction 
% Testing an active space
% But the true limits of an active space construction can only be be established by long experience.
}
\markchange{In order to investigate the basic properties of the AVAS procedure,} 
we now apply these criteria in calculations on various transition-metal complexes.

\subsection*{A. Ferrocene}

We begin by considering the electronic structure of ferrocene, \ce{Fe(C5H5)2}.
The ground state of ferrocene is dominated by a single configuration with Fe $d^6$. 
\markchange{Both the} MO analysis in Ref.~\onlinecite{fecp2_1} and our CI expansion coefficients indicate that the lowest excited states of ferrocene
have significant multiconfigurational character. 

We carried out an initial restricted  Hartree-Fock (ROHF) calculation for the singlet ground state.
We used the optimized $D_{5h}$ geometry  from Ref.~\onlinecite{fecp2_geom} (using the cc-pw-CVTZ basis set at the CCSD(T) level). 
In this geometry, the two cyclopentadienyl (Cp) rings are planar and the z-axis is aligned with the Cp-Fe-Cp axis.

\begin{figure}[t]
  \centering
  \includegraphics[width=.5\textwidth]{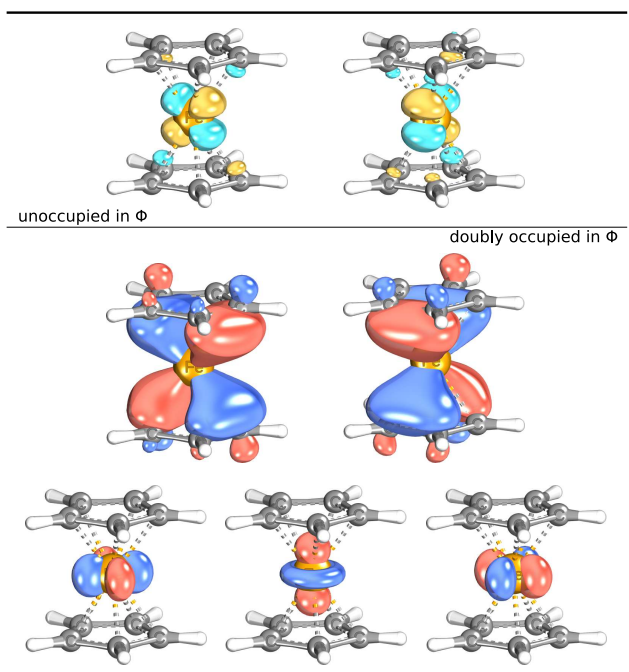}
  \caption{Ferrocene: The (10e,$\,$7o) AVAS initial active space for the five $3d$ orbitals of the iron atom. It is generated with threshold 0.1, based on the RHF wave function $\ket{\Phi}$ of the singlet ground state of the ferrocene molecule. After the AVAS transformation, all other occupied orbitals have exactly zero percent $d$-character, and all other unoccupied orbitals have 2.7\% \ce{Fe}~$3d$-character or less (see text).}
  \label{fig:FerroceneActiveSpace}
\end{figure}

As our target set of AOs, we first chose the five $3d$ orbitals of Fe.
Using a threshold of 0.1, the AVAS scheme produces a seven orbital active space: five orbitals from the occupied orbital space and two orbitals from the unoccupied orbital space \markchange{are combined into a \markchange{(10e,7o)} active space; the orbitals} are visualized in Fig.~\ref{fig:FerroceneActiveSpace}.
The five overlap eigenvalues above the threshold from the occupied space are 0.325, 0.325, 0.973, 0.973, 0.995, and the two eigenvalues from the unoccupied space are 0.675, 0.675.
There are three omitted unoccupied orbitals: two of them have only 2.7\% weight in the $3d$ AO space and $(3d_{xy}, 3d_{x^2-y^2})$ character, and one has 0.6\% weight in the $3d$ AO space, corresponding to the $3d_{z^2}$ AO. 
The two orbitals from the occupied orbital space with eigenvalues 0.325 and 0.325, as well as the two selected MOs from the unoccupied orbital space with eigenvalues 0.675, 0.675,
both have $(3d_{xz}, 3d_{yz})$ character; based on this, we conclude that $(3d_{xz}, 3d_{yz})$ are most strongly involved in bonding with the two Cp rings, in agreement with the bonding picture established in earlier  studies~\cite{fecp2_4,fecp2_1}.

To construct a second, larger active space, we next included the ten $p_z$ orbitals of the carbon atoms in the Cp rings into the target AO list. 
The AVAS construction then yields 15 orbitals above the 0.1 overlap threshold, giving an (18e,15o) active space: 9 of occupied character
and 6 of unoccupied character. 
\markchange{Of the 9 occupied orbitals, four do not carry Fe $3d$ character, instead belonging to the $\pi$-system of the Cp rings; two more have strong mixing with the Fe $(3d_{xz}, 3d_{yz})$ orbitals, and the final three represent almost-pure Fe $3d_{xy}$, $3d_{x^2-y^2}$, and $3d_{z^2}$ atomic orbitals (with $\geq95\%$ orbital weight on the Fe $3d$ AOs).
Among six unoccupied orbitals, two have no Fe $3d$ character, two have predominantly Fe $(3d_{xz}, 3d_{yz})$ character, and two have $<5\%$ character of Fe $3d_{xy}$ and $3d_{x^2-y^2}$ orbitals, respectively.
This agrees with our observations for the smaller (10e,7o) active space; namely, the additional orbitals in the (18e,15o) space are practically non-bonding in character.}
Lowering the \markchange{AVAS} threshold \markchange{from 0.1} to 0.05 gives a (22e,17o) active space\markchange{, which includes some further effectively non-bonding orbitals beyond the (18e,15o) active space}.

We calculated the ground state and singlet and triplet excited states of ferrocene with these active spaces at the CASCI and CASSCF levels. 
Previous studies\cite{fecp2_1} indicate that there are three low-lying $d \rightarrow d$ singlet transitions $(1 ^1E''_2, 1 ^1E''_1, 2 ^1E''_1)$ and
three low-lying $d \rightarrow d$ triplet transitions $(1 ^3E''_1, 1 ^3E''_2, 2 ^3E''_1)$.
These $d\rightarrow d$ transitions describe excitations from the three non-bonding orbitals (predominantly of $3d_{x^2-y^2}$, $3d_{xy}$ and $3d_{z^2}$ metal character,
as described above) to the two antibonding orbitals having mostly $(3d_{xz}, 3d_{yz})$ metal character. All
these excited states have multiconfigurational (but single-excitation) character. %% , arising in part from mixing between degenerate orbitals 
Note that the $E''_1$ and $E''_2$ states are doubly degenerate, thus there are 6 low-lying singlet and triplet excited states.
In the CASSCF calculations we therefore state-averaged over 7 roots and 6 roots in the singlet and triplet manifolds\markchange{,} respectively.

Table \ref{tab:tab1} displays the excitation energies, 
compared to  experimental and theoretical numbers from the literature, 
including singly excited configuration interaction (SECI) \cite{fecp2_2}, symmetry adapted cluster configuration interaction (SAC-CI) \cite{fecp2_1}
and time-dependent density functional theory (TD-DFT) calculations \cite{fecp2_3}.
Table \ref{tab:tab2}  compares the impact of using different active spaces with different methods.

The performance of CASCI for the smallest (10e,7o) active space is reasonable for most of the excited states, with the $2 ^1E''_1$ and $2 ^3E''_1$ states being exceptions. These two states have some Rydberg character,\cite{fecp2_1}
\markchange{and therefore} the valence CASCI overestimates these transitions by about 2 eV; \markchange{this effect can} also be seen in the errors of the SECI energies. 
% (Including the truncated orbitals with eigenvalues less than 0.1 into the active space 
%  changed the excitation energies by less than 0.01 eV). 
Averaging over all the states in the CASSCF seemed to spread the error over the states, lowering all the energies.
An accurate description of the differential correlation in the $2 ^1E''_1$ and $2 ^3E''_1$ states thus
requires a dynamic correlation treatment. We find excellent agreement for all states at the CASSCF+NEVPT2 level, 
with a largest error of only 0.21 eV. 

\markchange{Using the larger (18e,15o) active space, which includes the ligand $\pi$-orbitals, worsens the CASSCF excitation energies 
(except for the $2 ^1E''_1$ and $2 ^3E''_1$ states). However, incorporating dynamic 
correlation through NEVPT2 rebalances the states, improving CASCI and CASSCF energies (except for $1 ^1E''_2$ and $1 ^3E''_1$) and 
yielding better agreement with experiment and the (10e,7o) CASSCF+NEVPT2 excitation energies.
If we further go from the (18e,15o) active space to the (22e,17o) active space, by additionally including the truncated orbitals with eigenvalues below 0.1, the excitation energies change by less than 0.01 eV. 
Together, these observations indicate that the multiconfigurational character is already well 
converged in the smallest (10e,7o) active space.}

As discussed above, a second test of the quality of the AVAS active space is provided by the SVD decomposition 
of the overlap between the CASSCF-optimized active space and the \markchange{active space} initial guess. In the case of ideal coincidence, the SVD eigenvalues should be equal to 1.
The smallest SVD eigenvalues \markchange{we found were} 0.927 for the (10e,7o) active space and 0.906 for the (18e,15o) active space, respectively.
This indicates that \markchange{here} the AVAS provides a stable and accurate initial guess for the CASSCF procedure.

\begin{table}[h]
\centering
\begin{threeparttable}
\caption{Lowest singlet and triplet excitation energies of ferrocene: results from the (10e,7o) active space compared to other electronic structure methods. Energies are given in eV.}
\label{tab:tab1}
% \begin{center}\small
{\small
\begin{tabular}{  m{1 cm} | m{1.5 cm}| m{1.5 cm}| m{3 cm}|   m{1.6 cm} | m{1.5 cm} | m{1.7 cm} | m{2.2 cm }}
\hline \hline
State & CASCI     &CASSCF     & CASSCF+NEVPT2&  SAC-CI \cite{fecp2_1}& SECI  \cite{fecp2_2} & TD-DFT  \cite{fecp2_3} & Expt.* \\
         &(10e,7o) &(10e,7o)     & (10e,7o)  &                       &                      &PBE                    &   \\
\hline
$1 ^1E''_2$         & 2.85  & 2.10&2.79 & 2.11  & 2.63         &   2.90    & 2.8\cite{fecp2_e1},  2.7   \cite{fecp2_e2} \\                      
$1 ^1E''_1$         & 3.33  & 2.17&2.87 & 2.27  & 3.31         &   3.03    & 2.81\cite{fecp2_e1}, 2.98 \cite{fecp2_e2}\\
$2 ^1E''_1$         & 5.82  & 4.07&3.99 & 4.03   &5.74         &   3.60    & 3.82  \cite{fecp2_e1,fecp2_e2}\\
$1 ^3E''_1$         & 1.81  & 0.97&1.88 & 1.40  & 1.81-1.87 &              &  1.74\cite{fecp2_e1}   \\
$1 ^3E''_2$         & 1.84  & 1.07&2.03 & 1.68  & 1.81-1.87 &              &  2.05\cite{fecp2_e1}   \\
$2 ^3E''_1$         & 4.26  & 2.25&2.50 & 2.60 & 4.56       &              &  2.29-2.34\cite{fecp2_e1} \\
\hline \hline
\end{tabular}
% \end{center}
}
\begin{tablenotes}
\small
\item [*] From electronic absorption spectra of ferrocene in EPA glass, formed by cooling a EPA solution of ferrocene to liquid nitrogen temperature.
\end{tablenotes}
\end{threeparttable}
\end{table}

\begin{table}[h]
\centering
\begin{threeparttable}
\caption{Lowest singlet and triplet excitation energies of ferrocene: comparison of active spaces. Energies are given in eV.}
\label{tab:tab2}
{\small
\begin{tabular}{ c|cc|cc|cc|cc|l  }
\hline \hline
  &  \multicolumn{2}{c|}{ CASCI} & \multicolumn{2}{c|}{CASCI+NEVPT2} &  \multicolumn{2}{c|}{CASSCF}&  \multicolumn{2}{c|}{CASSCF+NEVPT2}&Expt.*\\
&(10e,7o) & (18e,15o)& (10e,7o) & (18e,15o)&(10e,7o) & (18e,15o)&(10e,7o) & (18e,15o)& \\
\hline
$1 ^1E''_2$  &2.85&3.51& 3.10&3.20   &2.10&2.88& 2.79&2.93& 2.8\cite{fecp2_e1},  2.7   \cite{fecp2_e2} \\
$1 ^1E''_1$  &3.33&3.90&  3.20&3.21   &2.17&3.45& 2.87&3.11& 2.81\cite{fecp2_e1}, 2.98 \cite{fecp2_e2}\\
$2 ^1E''_1$  &5.82&5.81&  4.24&4.00   &4.07&4.61& 3.99&4.06 & 3.82  \cite{fecp2_e1,fecp2_e2}\\
$1 ^3E''_1$  &1.81&2.46&  2.14&2.25   &0.97&1.91& 1.88&2.09&  1.74\cite{fecp2_e1}   \\
$1 ^3E''_2$  &1.84&2.48&  2.25&2.34   &1.07&2.37& 2.03&2.26&  2.05\cite{fecp2_e1}   \\
$2 ^3E''_1$  &4.26&4.33&  2.78&2.54   &2.24&2.64&  2.50&2.50 &  2.29-2.34\cite{fecp2_e1} \\
\hline \hline
\end{tabular}
}
\begin{tablenotes}
\small
% \item [*] From electronic absorption spectra of ferrocene in EPA glass, formed by cooling a EPA solution of ferrocene to liquid nitrogen temperature.
\item [*] See note in Tab.~\ref{tab:tab1}.
\end{tablenotes}
\end{threeparttable}
\end{table}

\subsection*{B. \ce{[Fe(NO)(CO)3]-}}

We next consider the complex anion \ce{[Fe(NO)(CO)3]-}, which exhibits catalytic activity in a range of organic reactions, and has been extensively characterized both theoretically and 
experimentally (see Ref.~\onlinecite{fenoco3} and references therein).
The complex features three-center bonds along both the \ce{Fe-N-O} axis and between \ce{Fe} and each pair of \ce{CO} ligands;
\cite{fenoco3} its catalytic mode of action exhibits a highly unusual nitrosyl-ligand based oxidation in some cases,\cite{klein2014fe}
and response to photo-activation in other cases.\cite{fenoco3_2}
Analysis of the ground-state CASSCF wavefunction and natural orbital occupations indicates that
it has some multiconfigurational character, and that it should be thought of as a Fe(0) species bound via two covalent 
$\pi$-bonds to the [NO$^{-}$].\cite{fenoco3}

\begin{figure}[t]
  \centering
  \includegraphics[width=.5\textwidth]{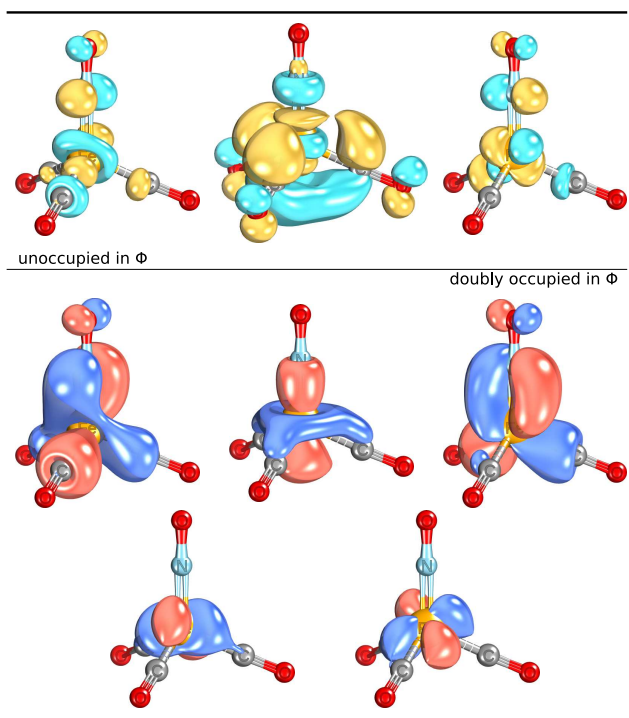}
  \caption{\ce{[Fe(NO)(CO)3]-}: The (10e,$\,$8o) AVAS initial active space for the five $3d$ orbitals of the iron atom, generated with threshold 0.1 from the ground state singlet RHF wave function.
  Two virtual orbials with 6\% $d$-character lie below the 10\% threshold and are omitted from the active space (see text); all other molecular orbitals are completely free of \ce{Fe}~$3d$-character after the AVAS transformation.}
  \label{fig:IronNitrosylActiveSpace}
\end{figure}

As in the previous example, we started with a RHF calculation for the singlet ground state\markchange{,} and 
for the simplest active space we chose five $3d$ orbitals of Fe as the target AOs. 
We used the geometry of Ref.~\onlinecite{fenoco3_2}. 
Using an overlap threshold of 0.1 gives rise to five occupied orbitals and three unoccupied orbitals for the active space; two unoccupied orbitals with 
only 6\% weight in the $3d$ orbital space lie below the threshold for active space inclusion.
Of the three unoccupied orbitals included in the active space, two have $(3d_{yz},3d_{xy})$ and $(3d_{xz},3d_{x^2-y^2})$ character, respectively, while the third one has mostly $3d_{z^2}$ character.
The resulting active space is visualized in Fig.~\ref{fig:IronNitrosylActiveSpace}.
Additionally including the nitrogen $2p$ orbitals \markchange{into the set of target AOs (with the same threshold)} gives a (16e,14o) \markchange{active space; this adds}
three MOs with $2p$ character, involved in the three-center \ce{Fe-N-O} bonds, and three involved in \ce{N-Fe-CO} type bonds.

Unfortunately, there is no gas phase experimental excitation data for this system. However, theoretical vertical excitation energies
from state-averaged CASSCF calculations followed by MRCI+Q, using the def2-TZVPP basis set (omitting $g$-functions) have previously been reported,\cite{fenoco3_2} which we can compare against.
% (as well as PBE density functional data) 
We used CASSCF and NEVPT2 to compute the vertical transition energies
averaging over five singlet and four triplet states, as in Ref.~\onlinecite{fenoco3_2}.
The smallest SVD eigenvalue for the active space overlap with the initial guess for the (10e,8o) active space is 0.806, indicating
that AVAS provides a good initial guess. For the (16e,14o) active space, the lowest SVD eigenvalue decreases to 0.652---apparently adding only the nitrogen $2p$ orbitals,
without also adding the carbon $2p$ orbitals, leads to a less balanced active space compared to the (10e,8o) AVAS initial active space.

The vertical excitation energies from CASSCF with the (10e,8o) active space are in better agreement with the CASSCF/MRCI+Q excitation 
energies than with CASSCF results from Ref.~\onlinecite{fenoco3_2}. This indicates that the (10e,8o) active space constructed using AVAS 
provides a more balanced decription of electron correlation than the larger active spaces.

However, the NEVPT2 dynamical correlation treatment significantly raises the obtained excitations energies above the MRCI+Q values.
Similarly, CASSCF calculations with the larger (16e,14o) active space also yield significantly higher excitation
energies than with the (10e,8o) active space, and the CASSCF+NEVPT2 excitation energies in this larger space are also
fairly different from the values obtained from the (10e,8o) active space. This lack of stability with respect to the active space size indicates
that the excited states are not benign electronically, and that their accurate description requires a more sophisticated dynamic correlation treatment beyond 2nd order perturbation theory.
This is supported by the MRCI+Q study in Ref.~\onlinecite{fenoco3_2}, where the (empirical) Q-contribution to the excitation energy is as large as 0.2 eV.

To further substantiate these claims, we also computed CASSCF+NEVPT2 results for the same manually selected (14e,9o) initial active space as described
in Ref.~\onlinecite{fenoco3_2} (in these calculations, the initial 14 active orbitals were manually selected for Fe $d$ and NO $\pi$ and $\pi^*$-character by visual inspection of KS-DFT/PBE
orbitals computed with the def2-TZVPP basis set; the CASSCF excitation energies thus obtained reproduced the results reported in the supporting information of Ref.~\onlinecite{fenoco3_2} with better than 0.01~eV accuracy).
By comparing the results of (our) NEVPT2 and (the referenced) MRCI+Q for the same active space of Ref.~\onlinecite{fenoco3_2}, we can separate the effect of the dynamic correlation treatment from the quality of the active space.
We see that CASSCF+NEVPT2 calculations performed with our automatically constructed (10e,8o) active space and the manually selected (14e,9o) active space
show a fair agreement in the case of singlet excited states; however, the CASSCF+NEVPT2 method overestimates the energies of the triplet excited states with the (14e,9o) active space
by 0.5--0.75 eV more than with the (10e,8o) active space, compared to the CASSCF/MRCI+Q transition energies.
Combined, these facts strongly suggest that the approximate NEVPT2 correlation treatment is the primary cause of deviation from the
MRCI+Q reference values, rather \markchange{than} our automatically constructed active space, and that the smaller AVAS provides a more balanced description of this system.

\begin{table}[h]
\centering
{\small
\begin{threeparttable}
\caption{Vertical excitation energies from the ground state to the lowest singlet and triplet states of \ce{[Fe(NO)(CO)3]-}. Energies are given in eV.}
\label{tab:tab3}
 \begin{tabular}{ c|cc|cc|c|c|c}
\hline \hline
 State &  \multicolumn{2}{c|}{ CASSCF} & \multicolumn{2}{c|}{CASSCF+NEVPT2} &  CASSCF\cite{fenoco3_2} &CASSCF/MRCI+Q \cite{fenoco3_2}&CASSCF+NEVPT2 \\
& (10e,8o) & (16e,14o) & (10e,8o) &(16e,14o) & (14e,9o)*& (14e,9o)*& (14e,9o)* \\
\hline
$^1A_2$ & 3.14 & 4.37 &4.04 & 3.76&2.78 &3.26 &4.13 \\
$^1E$   & 3.58 & 4.86 &4.29 & 3.98 & 3.18&3.53 &4.29\\
$^1A_1$ & 3.54 & 5.17 &4.33 & 4.25 & 3.22&3.64 &4.40\\
$^3A_1$ & 2.27 & 2.44 &2.63 & 2.43 &1.76&2.32 &3.40\\
$^3E$   & 2.82 & 3.41 &3.51 & 3.20 	&2.44&2.96&4.02\\
$^3A_2$ & 2.89 & 4.04 &3.80 & 3.82 &2.56&3.15&4.39 \\
\hline \hline
\end{tabular}
\begin{tablenotes}
\small
\item [*] The (14e,9o) active space consisted of Fe $d$ orbitals and the NO $\pi$ and $\pi*$ \markchange{(manually chosen according to Ref.~\onlinecite{fenoco3_2}).}
\end{tablenotes}
\end{threeparttable}
}
\end{table}

\subsection*{C. \ce{FeO4^2-} }

As our next system, we consider the bare tetraoxoferrate (VI) ion, \ce{FeO4^2-}.
We assume a tetrahedral \ce{FeO4^2-} cluster with  an Fe--O distance of 1.660\,{\AA}. 

We started with a ROHF calculation for the $^3A_2$ ground state. 
Including only $3d$ orbitals into the target AO set gives a (8e,8o) active space.
Three unoccupied MOs have 34.5\% weight in the $3d$ orbital space, with $3d_{yz}$, $3d_{xz}$, and $3d_{x^2-y^2}$ character,
but are mostly centred on the ligands; the occupied MOs have mostly metal character.
This indicates that some of the low-lying excitations are charge-transfer excitations.

An earlier study~\cite{feo4m2} found that the ground and excited states cannot be described by
a simple Ligand Field Theory $d^2$ model\markchange{; rather, they} contain superpositions of a large number of configurations, including  
ligand-to-metal excitations\markchange{.} From this it has been argued that it is insufficient
to only  consider  molecular orbitals with Fe $3d$ character in
the active space to describe excited states. 
Indeed, we find \markchange{that} CASSCF+NEVPT2 calculations with the (8e,8o) active space (generated only with the $3d$ orbitals in the target AO set)
significantly overestimate the excited states \markchange{energies}, by about $\approx$ 6500 cm$^{-1}$  ($\approx$0.81~eV), compared to experiment.

For this reason, we expanded the target AO list to the five 3$d$ orbitals of Fe and 2$p$ orbitals of all four O atoms.
Using option \ref{onlyalpha} to transform the alpha orbitals, our scheme with the 0.1 overlap threshold produced 14 
occupied orbitals and 3 unoccupied orbitals, resulting in a (26e,17o) active space. 

We calculated the vertical excitation energies of \ce{FeO4^2-}, namely 
transition energies from the ground $^3A_2$ state to the first two excited states, $^1E$ and $^1A_1$ (see Table \ref{tab:tab4}),
using CASCI and CASSCF (state-averaged over three singlet states and one triplet state) and CASCI+NEVPT2,
comparing to previously reported RASSCF and experimental numbers.
The CASCI calculations significantly overestimate the excitation energies\markchange{;} however, this is significantly improved
by optimizing the orbitals using state-averaged CASSCF.
The smallest SVD singular value for the active space overlap between the initial
and optimized active orbitals is 0.968, indicating that
our initial active orbitals provide a very good guess for the CASSCF procedure and only require
a little relaxation to yield good agrement with experiment. 

Including dynamic correlation by means of NEVPT2 on top of CASCI or CASSCF significantly improves the results.
Note that the difference between our CASSCF excitation energies and those in Ref.~\onlinecite{feo4m2}  with 17 orbitals, obtained with 
the RASSCF method\markchange{,} reflects both the slightly different basis \markchange{set} as well as \markchange{the} truncated CI \markchange{configuration} space in RAS.

\begin{table}
{\small
\centering
\begin{threeparttable}
\caption{Calculated and experimental low-lying excitation energies of  \ce{FeO4^2-}. Energies are given in cm$^{-1}$.}
\label{tab:tab4}
\begin{tabular}{ c| c| c |c |c|c | c}
\hline \hline
State &  CASCI     & CASSCF & CASCI+NEVPT2&CASSCF+NEVPT2&RASSCF \cite{feo4m2} & Expt.* \\
         & (26e,17o)  &(26e,17o) &(26e,17o) &(26e,17o) &  (26,4,0;12,5,0)               &  \\
\hline
$^1E$            & 9882, 9899  &  7462	  & 6252, 6268& 6548, 6550&6300    & 6209, 6219  \cite{feo4m2_e1}   \\
                      &          &            &    &          & & 6219, 6230   \cite{feo4m2_e2}  \\
$^1A_1$        &14703  & 10710  &9006& 9471     &9200    & 9119 \cite{feo4m2_e1}, 9176   \cite{feo4m2_e2} \\         
\hline \hline
\end{tabular}
\begin{tablenotes}
\small
\item [*] From single-crystal polarized electronic absorption spectra.
\end{tablenotes}
\end{threeparttable}
}
\end{table}

\subsection*{D. \ce{VOCl4^2-} }	
\label{sec:vocl4}

We now consider the oxotetrachlorovanadate(IV) anion, \ce{VOCl4^2-}.
We use a square pyramidal geometry for \ce{VOCl4^2-}\markchange{,} as in Ref.~\onlinecite{Vancoillie} (although 
we use a different orientation: the V atom is at the origin, the O atom is on the z axis above the x-y plane and the Cl atoms are below the x-y plane). 

In this complex, vanadium is in a $d^1$ configuration. 
As in the next example, here the $d\to d$ excitations  do not have much multiconfigurational character. However, it is important
that multireference methods (and their active spaces) provide a balanced description of all states, not just multiconfigurational
ones. The vanadium complex provides a system to test this in an early transition metal (single-reference) problem.

If we choose a set of five AOs, representing the five $3d$ orbitals of a vanadium atom, 
we obtain five occupied and four unoccupied MOs from the ROHF reference wavefunction 
using the AO-projector option \ref{onlyalpha}. One of the occupied MOs 
is a non-bonding $3d_{xy}$ atomic orbital, while the
other V $3d$ \markchange{AOs} strongly mix with the valence orbitals of the oxygen atom and four chlorine atoms. 
This results in four doubly occupied bonding MOs and four anti-bonding MOs which are unoccupied in the ground state. 
The two unoccupied MOs have 69.4\% overlap with the $3d$ orbital space and carry $(3d_{xz},3d_{yz})$ character,
other two have 54.0\% and 70.3\% overlap with the $3d$ orbital space and have $(3d_{z^2}$ and $3d_{x^2-y^2})$ character, respectively;
all four unoccupied MOs have mostly metal character.

Using the (9e,9o) active space we calculated the lowest transitions, which are essentially $d \rightarrow d$ in nature.
There are four possible ligand-field transitions from the highest non-bonding $3d_{xy}$ orbital to four unoccupied MOs, thus
in CASSCF we averaged over five doublet states. Table \ref{tab:tab5} summarizes the low-energy vertical $d \rightarrow d$ energies.
The CASSCF method with the small (9e,9o) active space gives an accurate $^2B_1$ state, but strongly overestimates 
the other excited states. 
Using NEVPT2 to treat the dynamic correlation on top of CASSCF significantly improves the excited \markchange{state energies},
resulting in a good agreement with the experimental values and \markchange{with} CASPT2 results  obtained with \markchange{a} (11e,10o) space~\cite{Vancoillie}.
\markchange{This} (11e,11o) space is similar to ours\markchange{,} with the addition of the oxygen $2p$ shell.
 
We also construct a larger (33e,21o) active space, including the $2p$ orbitals of O and the $3p$ orbitals of Cl into the target
AO list. In this larger space, to reduce computational cost, we used CASCI+NEVPT2 rather than CASSCF+NEVPT2.
The excited states from CASCI+NEVPT2 with the (33e,21o) active space are also in excellent agreement with the experimental data.
The stability of the CASCI/CASSCF+NEVPT2 excitations with respect to expanding the active space \markchange{confirms} that the correlation
is well converged  by all these treatments.
 
In the CASSCF calculation with the (9e,9o) active space, the smallest SVD eigenvalue for the active space overlap \markchange{between the initial and the optimized active orbitals amounts to} 0.821. This implies that, in this case, AVAS provides a reasonable, but not perfect\markchange{, initial} guess for CASSCF. 

We also used this complex to test and compare options \ref{onlyalpha} and \ref{openshell}, as described in Sec.~\ref{sec:openshell}, for constructing the active space with a ROHF reference determinant $\ket{\Phi}$.
The excitation energies, calculated with these two options, differ by less than 100~cm$^{-1}$ ($\approx$0.01~eV). However, \ce{VOCl4^2-}'s ground state has only one singly occupied orbital,
and it is possible that larger differences will occur for systems with ground states of higher spin.

\begin{table}
\begin{threeparttable}
\caption{Vertical excitation energies to the lowest doublet excited states from the $^2B_2$ state ground of  \ce{VOCl4^2-}. Energies are given in cm$^{-1}$.  }
\label{tab:tab5}
\begin{tabular}{  c|c|c|c|c|c}
\hline \hline
State & CASSCF& CASSCF+NEVPT2     &  CASCI+NEVPT2    &   CASPT2\cite{Vancoillie} & Exp.*  \\
       &(9e,9o)  & (9e,9o)       & (33e,21o)               & (11e,10o)  & \\
\hline
$^2B_1$    &11736 & 13044  &  12123 & 12667 &  11600\cite{vocl4_2}    \\
$^2E$      &18650 & 14537  &  13722 & 13620  &  13700\cite{vocl4_2}      \\
$^2A_1$    &33808 & 28586  &  26824 &        &     \\             
\hline \hline
\end{tabular}
\begin{tablenotes}
\small
\item [*] From single-crystal polarized electronic absorption spectrum
\end{tablenotes}
\end{threeparttable}
\end{table}

\subsection*{E. \ce{[CuCl4]^2-}  }	

We finally consider the  $D_{4h}$ \ce{[CuCl4]^2-} complex, with a Cu--Cl bond length of  2.291 {\AA}, as in Ref.~\onlinecite{Vancoillie}.
As in the vanadium system, the $d\to d$ transitions are single-reference in character: this complex provides a late transition metal example.

Using the $3d$ AOs of Cu as the target AOs and a default cutoff of 0.1,  we obtain only 5 occupied MOs and no unoccupied MOs\markchange{. This result might surprise} at first glance. However, \markchange{in this case,} the antibonding orbitals have ligand character,
and thus there are no unoccupied MOs having more than 5\% $3d$ character. 
The lowest ligand-field transitions arise from the excitation of electrons from the doubly-occupied MOs 
with dominant $3d$ character to the singly occupied MOs with $3d$ character.

Despite the fact that the lowest transitions happen mostly \emph{within} the $3d$ orbital space,
such a small (9e, 5o) active space is insufficient to describe them\markchange{\ at the CASSCF level---because}
the nearly filled space leaves no room for electron correlation. 
CASSCF+NEVPT2 however, provides good agreement with the experimental numbers (see Table \ref{tab:tab6}).
To see the effect of a larger active space, we also included the
 $3p$ AOs of Cl in the target AO list, obtaining a (33e,17o) active space. The corresponding CASSCF
excitation energies are still poor, indicating that the necessary correlation is not of valence character.
The CASSCF+NEVPT2 excitation energies in this larger space, however, remain in very good agreement with experiment.
The insensitivity to active space indicates that correlations are well converged in the CASSCF+NEVPT2 treatment.

\markchange{In CASSCF calculations we averaged over five doublet states, each having four doubly occupied $d$ orbitals and one $d$-orbital with a single electron.}
The smallest SVD eigenvalue for the active space overlap with the initial guess in converged CASSCF calculations
is equal to 0.930 in the case of the small (9e,5o) active space and 0.985 for the (33e,17o) active space, indicating
that AVAS provides a good initial guess for CASSCF. %% However inclusion of dynamic correlation is required for the accurate description of excited states. 

\begin{table}[h]
\begin{threeparttable}
\caption{Low-lying excitation energies of \ce{[CuCl4]^2-}.  Energies are given in cm$^{-1}$.}
\label{tab:tab6}
\begin{tabular}{ c|cc|cc|c|cc}
\hline \hline
 State &  \multicolumn{2}{c|}{ CASSCF} & \multicolumn{2}{c|}{CASSCF+NEVPT2}  & CASPT2  \cite{Vancoillie}   &   Exp.* \\
&(9e,5o)  &  (33e,17o)&(9e,5o)  &  (33e,17o)& (11e,11o)& \\
\hline
$1 ^2B_{2g}$     &6588&6588&10675  &10666& 11321     & 10500 \cite{Willet}   \\
$1 ^2E_{g}$        &8727&8728&12832 &12947& 13379      &12800 \cite{Willet} \\
$1 ^2A_{1g}$      &9690&9692&14021  &14135& 14597      &\\
\hline \hline
\end{tabular}
\begin{tablenotes}
\small
\item [*] From polarized absorption spectrum of a single-crystal $D_{4h}$ \ce{[CuCl4]^2-}
\end{tablenotes}
\end{threeparttable}
\end{table}

\subsection*{F. Fenton reaction: an example of homolytic bond dissociation}\label{sec:FentonReaction}
{\markchangesection
When modeling chemical reactions with multi-configuration methods, one particularly challenging problem is the selection of active spaces capable of representing all relevant configurations in a homolytic bond dissociation process.
In fact, aside from special cases, where e.g. a reasonable active space choice along the reaction path is enabled by ``accidental'' factors, such as molecular symmetry, or all the relevant orbitals are energetically well-separated from and unmixed with other orbitals, or the molecule is simply small enough to allow for a full-valence active space treatment, active space selection constitutes
the primary bottleneck in the real-world applications of multi-configurational methods to reaction chemistry.

%% In fact, we are not aware of any generic procedure for creating such spaces in general molecules\footnote{by ``in general molecules'' we mean molecules in which }, apart from the recently introduced technique of Stein and coworkers\cite{stein:ActiveSpaceSelectionForReactions} {(\color{blue}!!FIXME: wasn't there also a paper in JCTC 2017 on this? With DMRG? and joining active spaces on all frames of a reaction? I could swear I read one of those... I thought this was from Reiher's group)}.
%% Our understanding is that, at this moment, this active space selection problem effectively precludes the ab-initio modeling of homolytic bond dissociation processes in most real-world application scenarios, unless molecular symmetry or other specific factors simplify the process.

However, as rationalized in Sec.~\ref{sec:WhatIsAnActiveSpace}, we expect the AVAS scheme to be applicable when studying homolytic bond dissociation processes. 
To illustrate this capability, we here consider one key step of a Fenton reaction. Concretely, we consider the homolytic splitting of hydrogen peroxide (\ce{H2O2}) by aqueous ferrous (Fe(II)) iron at low pH, to yield \ce{OH^{.}} radicals and an aqueous ferric (Fe(III)) species. Summarized, we examine the inner step of
% \ce{[Fe^{2+}(aq)-HOOH] -> [Fe^{3+}(aq)-OH^{-}] + OH^{.}}.
\begin{center}
\ce{Fe^{2+} + H2O2 -> [Fe^{2+}(HOOH)] -> [Fe^{3+}(OH^{-})] + OH^{.} -> Fe^{3+} + OH^{.} + OH^{-}}.
\end{center}
Variants of this reaction have been proposed as possibly relevant elementary steps in biochemical processes involving iron-oxo (\ce{Fe(IV)=O}) species (including such as performed by cytochromes P450\cite{dunford:Iron23VsHydrogenPeroxideReview}), but this chemistry has been the subject of intense mechanistic debate since its inception in 1894\cite{fenton:FentonReaction1894}.
%(including such as occuring in the cytochrome P450 catalytic cycle)
The history, background, and significance of this reaction, as well as the surrounding controversies, are discussed in Refs.~\onlinecite{dunford:Iron23VsHydrogenPeroxideReview}, \onlinecite{kremer:fenton1999}, and \onlinecite{petit:FentonReaction}.

Here no attempt is made at quantitatively describing the reaction, or at resolving any of the controversies it involves; rather, we only examine whether AVAS is capable of producing an active space capable of \emph{qualitatively} describing the reaction mechanism of a non-trivial homolytic bond dissociation along the entire IRC. In particular, no solvation effects are considered.

\begin{figure}
   \centering
   \includegraphics[width=0.9\columnwidth]{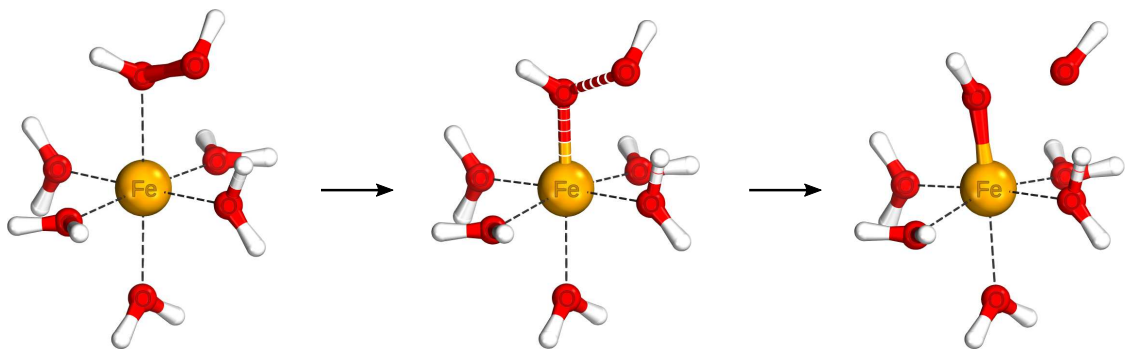}
   \caption{Structural model for the homolytic bond dissociation process of the Fenton reaction: a process of homolytic bond dissociation of hydrogen peroxide \ce{H2O2}, catalyzed by aqueous iron(II). The \ce{[Fe(H2O)5(H2O2)]^{2+}} complex (left) is formed when \ce{H2O2} enters the coordination sphere of an octahedral \ce{[Fe(H2O)6]^{2+}} complex, a model complex for the aqueous \ce{Fe^{2+}(aq)}. The homolytic bond cleavage of the coordinated hydrogen peroxide in the \ce{Fe(H2O)5(H2O2)^{2+}} complex leads to the formation of two spin-coupled radical fragments: a neutral hydroxide radical \ce{OH^{.}} and an iron(III) complex \ce{[Fe(OH)(H2O)_5]^{2+}}.
   The center picture depicts the transition state (UB3LYP/def2-TZVP), the left and the right pictures depict two structures along the computed IRC on the reactant and product side of the reaction model.}
   \label{fig:FentonGeometry}
\end{figure}

% \markchange{(
% )}

To model the Fenton reaction, we considered the \ce{[Fe(H2O)5(H2O2)]^{2+}} complex as an initial reagent. 
It represents an octahedral \ce{[Fe(H2O)6]^2+} complex, a model for the aqueous ferrous ion, 
with one water molecule substituted by the hydrogen peroxide molecule. 
Such a complex is expected to be formed when \ce{H2O2} enters the coordination sphere of \ce{[Fe(H2O)6]^2+}.
The homolytic bond cleavage/dissociation of the \ce{[Fe(H2O)5(H2O2)]^{2+}} complex leads to the formation of two spin-coupled radical fragments: ferric \ce{[Fe(H2O)5(OH)]^{2+}} and hydroxyl \ce{OH} radicals \cite{petit:FentonReaction}. 
For this process we computed a reaction path along the intrinsic reaction coordinate, at the level of UB3LYP/def2-TZVP as described in Sec.~\ref{sec:ComputationalDetails} (the geometry optimizations were started with structures from Ref.~\onlinecite{petit:FentonReaction}).
The reaction model is visualized in Fig.~\ref{fig:FentonGeometry}.
The reactant complex has four unpaired electrons on iron and a total spin quantum number of $S=4/2$.

To build the active space, we first computed ROHF wave functions (with four unpaired electrons) for each structure along the IRC.
To ensure the convergence for the ROHF method and retain a continuous character of the ROHF solution along the entire reaction path, we used the orbitals obtained for the previous geometry as an initial guess for the next one, starting at the reactant side.
The actual active space was then formed by the AVAS procedure, in which we used the three $2p$ orbitals of two dissociating oxygen atoms and five $3d$ orbitals of \ce{Fe} as target AOs.
For the open-shell treatment, we invoked option 3 from Sec.~\ref{sec:openshell}; that is, we applied the AVAS projection only on doubly-occupied and virtual orbitals from ROHF wavefunction, and the four singly-occupied $3d$
orbitals of Fe were added unchanged to the active space to keep the correct spin $S$ number for the entire system.
For the initial complex, AVAS with a 10\% threshold produced 12 occupied orbitals and 1 unoccupied orbital to form the (20e,13o) active space.
To retain consistency of the active space along the reaction path, we then fixed the size of the active space to the values thus obtained; that is, for each geometry we chose the 12 occupied orbitals corresponding to the 12 largest overlap eigenvalues with the target space, and the one virtual orbital with the largest overlap with the target space.

\begin{figure}
   \centering
   \includegraphics[width=.9\columnwidth]{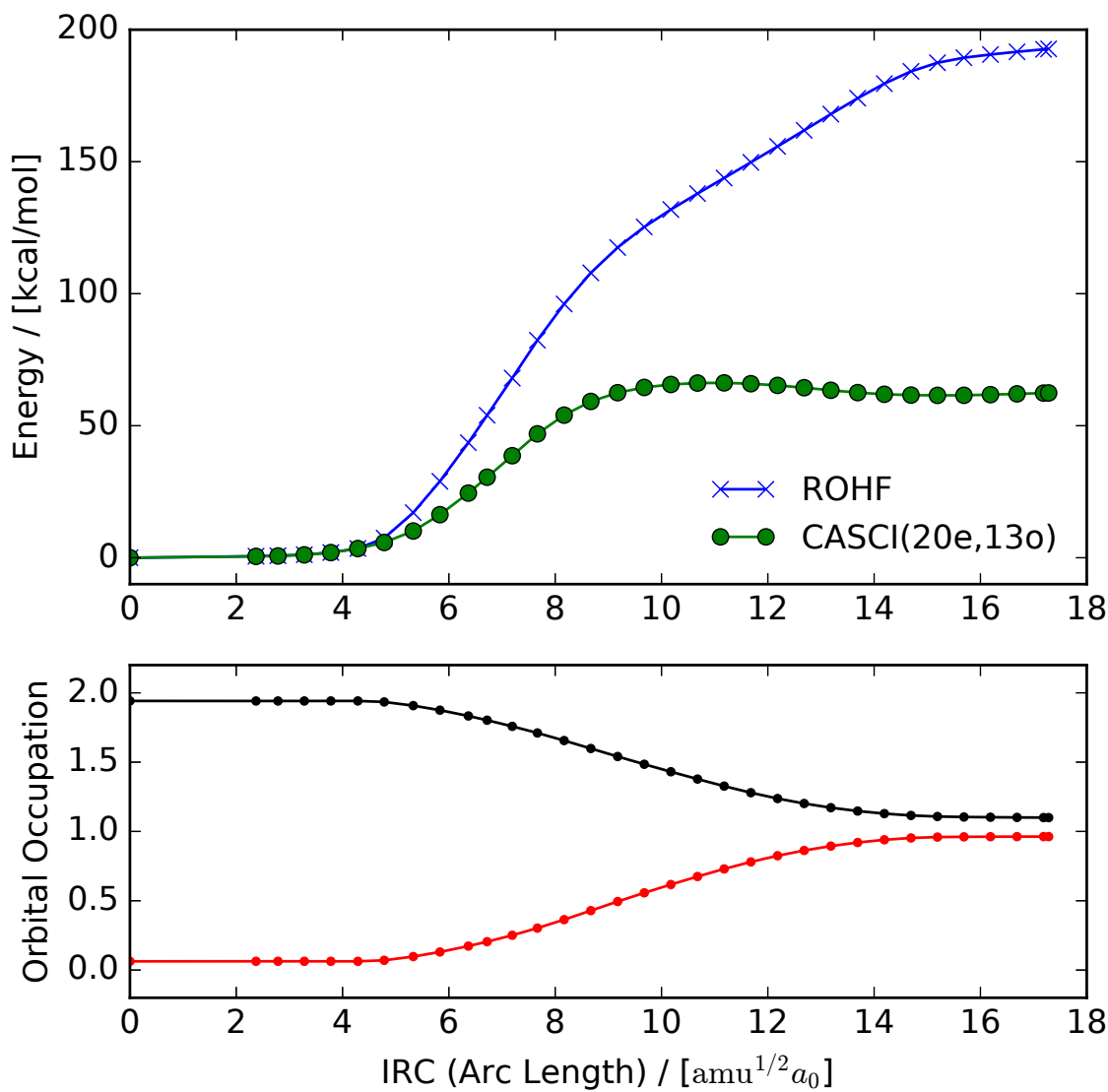}
   \caption{Computed data along the intrinsic reaction coordinate of the homolytic bond dissociation process visualized in Fig.~\ref{fig:FentonGeometry}. cc-pVTZ basis sets were employed,
   and CASCI was executed with DMRG-CI (M=1600), and the (20e,13o)-active space (from AVAS with the two \ce{O} $2p$ electron shells and the \ce{Fe} $3d$ electron shell as target AOs; see text).
   Top: ROHF and CASCI energies. Bottom: occupation numbers of the two active orbitals in the cleaved \ce{O-O} bond (all other active orbitals retain an orbital configuration of $\approx 1.0$ (for the Fe $3d$ active orbitals) and $\approx 2.0$ (for the other active orbitals) along the entire IRC).
   }
   \label{fig:FentonEnergyFigure}
\end{figure}

With this active space, multi-configuration calculations along the reaction path were executed.
We wish to probe the capability of the AVAS to describe the qualitative features of the electronic structure  on \emph{both} sides of the reaction, despite the fact that AVAS was constructed from a ROHF wave function, which is itself incorrect on the product side. For this reason, the orbitals \emph{were not optimized} in the multiconfiguration treatment---that is, we use (approximate) CASCI with ROHF orbitals as a multi-configuration method, and not (orbital-optimized) CASSCF.
Concretely, the CASCI energies were approximated with DMRG-CI (M=1600), which should be essentially exact with this active space.

The obtained energies are displayed in Fig.~\ref{fig:FentonEnergyFigure} (top).
During the process, the ROHF energy rises, because a ROHF determinant is incapable of qualitatively describing the electronic structure of the product complex.
In contrast, with the same ROHF orbitals, the computed CASCI energies show an energy maximum, and a decreasing energy towards the product side.
In Fig.~\ref{fig:FentonEnergyFigure} (bottom) we also plot the CASCI-occupation numbers of two active MOs participating in this homolytic bond cleavage along the IRC.
As expected for a homolytic bond cleavage, in which one doubly occupied molecular orbital splits into two singly occupied molecular orbitals, the occupation numbers go from $\approx$2.0/0.0 to $\approx$1.0/1.0 during the process.
This confirms the formation of two radicals on the product side.
Concretely, the active orbital obtained from the virtual side of the AVAS construction at ROHF level changes its occupation
number from 0.06 to 0.96 along the IRC, at the DMRG-CI level.

As expected at this level of theory, the absolute energetics of the process are not well reproduced (compared to accurate CCSD(T)-F12 calculations\cite{petit:FentonReaction}).
However, we conclude that the qualitative features of the process are correctly described with the given active space.
In particular, the AVAS contains the relevant active orbitals for describing the electronic structure of both the reactant side and product side of the reaction---even though the AVAS is constructed from a ROHF wave function which is qualitatively incorrect on the product side.
This observation is consistent with the earlier finding in DMET that relevant entanglement spaces can be constructed from Hartree-Fock wave functions even in cases where the latter are qualitatively incorrect.\cite{GK_GKC2}
All geometries, energies and occupation numbers are available in Supporting Information.
}

\section{\markchange{Speculative extensions for complex cases}}\label{sec:BeyondAvas}
{\markchangesection
We believe that AVAS is useful for the application cases fulfilling the premises described in Sec.~\ref{sec:WhatIsAnActiveSpace}.
However, it is clear that various application scenarios fall outside of this range.
We here outline two straight-forward extensions of the AVAS concept, which may extend its range of applicability.
% The details and practical performance of these approaches will be investigated elsewhere.
Investigations of the details and practical performance of these approaches are beyond the scope of this work and will be done elsewhere.

% Details of these approaches will be investigated elsewhere.

\subsection{Active spaces beyond the valence space}\label{sec:BeyondTheValenceSpace}
For some combinations of electronic structure methods and/or application scenarios, it may be desirable to construct active spaces capable of explicitly representing electronic degrees of freedom which cannot be represented within a (minimal) valence-AO basis.
One such case is the treatment of Rydberg excited states: describing such states on even a qualitative level will require diffuse orbitals in the active space.

The other, more common, case, is the extension of the active space in order to improve the \emph{quantitative} accuracy of computations in certain cases.
For example, the accuracy of CASPT2 (and related PT2 methods) for transition metal complexes is often substantialy improved if a correlating second radial shell of $d$-AOs is included into the active space. 
This is known as the ``double-shell effect'',\cite{Dunning:DoubleShell1,Dunning:DoubleShell2,Pierloot:NonDynCorrTMC} and rests on the fact that doing so, effectively, relocates the treatment of $d$-electron radial electron correlation from the (very approximate) ``PT2 part'' of the method to the (very accurate) ``CAS part''.

While in this work only valence-spanned active spaces are considered, the AVAS procedure should be able to also create such extended active spaces---by including the additional desired non-valence AOs in the target AO space. This can be done by simply taking any non-minimal atomic natural orbital basis set (such as ANO-RCC,\cite{widmark:AnoRccHHe,roos:AnoRccMainGroups} ano-pVnZ,\cite{neese:ano2011} ANO-VT-XZ,\cite{ANO_VT} or any other) as source for these target AOs---rather than the MINAO basis used here.
For example, a ``double-shell'' active space for a 3$d$ transition metal atom could be created by including \emph{both} 3$d$ and 4$d$ AOs in the ``target'' AO space of AVAS. One could even add an additional shell of tight $f$ functions, to follow the idea of ``correlation consistency''\cite{dunning:ccpvnz1,almlof:AnoCorrelationConsistency1,almlof:AnoCorrelationConsistency2} of dynamic correlation---although doing so is not common practice.
% Details of this approach will be investigated elsewhere.

\subsection{Handling complex metal/ligand interactions or multi-nuclear coordination complexes}
We expect that normally the simple $d$-electron AVAS will be sufficient to qualitatively represent the multi-reference wave functions of ground states and (at least) $d\mapsto d$-excited states of single-metal coordination complexes. However, there are many application scenarios in which either complicated and non-transparent metal-ligand interactions, or the presence of multiple metal cores in coordination complexes, would make a straightforward active space construction via AVAS either questionable (if ligand AOs or additional metal AOs are not included in the target space) or infeasible (if such additional AOs are included, but the so-created active space becomes too large for convenient quantitative handling).

In the presence of such effects it may be warranted to \emph{combine} AVAS with either one of several methods of constructing approximate FCI wave functions in large active spaces (\emph{vide infra}), or with the entanglement-based active-space construction procedure of Reiher and coworkers\cite{Keller:ActiveSpaceSelection,stein:ActiveSpaceSelectionForReactions}.

In the simplest case of suspected complex metal-ligand interactions in a single-metal complex, one might, for example, first create an AVAS by including as target AOs both the $d$-electrons of the transition metal, as well as \emph{all} ligand AOs suspected to play an important role (e.g., all ligand valence AOs from the first coordination sphere, or all $p_z$-orbtials of a large $\pi$-system coordinated to the metal atom).
In practice, the active space created from this choice will frequently be too large for an accurate quantitative multi-reference calculation including dynamic correlation. 
However, there are methods capable of providing qualitative wave functions and first-order reduced density matrices at an approximate CASCI level for such active spaces.
For large active spaces without too many open shells, methods such as SHCI~\cite{sharma:shci,holmes:hci} or FCIQMC~\cite{booth:FciQmcOriginal,cleland:FciQmcInitiatorApprox,booth:FciQmcExcitedStates1,overy:FciQmcRdm} can
provide an efficient description, while for active spaces with a larger number of open shells, but fewer than 50 orbitals, QC-DMRG~\cite{white:qcdmrg,mitrushenkov:qcdmrg1,mitrushenkov:qcdmrg2,chan:qcdmrg1,legeza:qcdmrgaccuracy,chan:qcdmrg2,reiher:dmrgOrbitalOrdering1,chan:DmrgWithNonOrthOrbitals,reiher:DmrgEnvironmentStates,chan:qcdmrgReview2011,sharma:QcdmrgSpinAdapted,chan:qcdmrgReview2015} is a reliable approach.
Once such a qualitative wave function is present, it can be used to determine which orbitals are \emph{not needed} in the active space.

%% However, for 
%% active spaces without too many open shells
%% active spaces up to 30--50 orbitals, methods such as QC-DMRG,\cite{white:qcdmrg,mitrushenkov:qcdmrg1,mitrushenkov:qcdmrg2,chan:qcdmrg1,legeza:qcdmrgaccuracy,chan:qcdmrg2,reiher:dmrgOrbitalOrdering1,chan:DmrgWithNonOrthOrbitals,reiher:DmrgEnvironmentStates,chan:qcdmrgReview2011,sharma:QcdmrgSpinAdapted,chan:qcdmrgReview2015} FCIQMC,\cite{booth:FciQmcOriginal,cleland:FciQmcInitiatorApprox,booth:FciQmcExcitedStates1,overy:FciQmcRdm} and, in particular, the extremely powerful SHCI\cite{sharma:shci,holmes:hci} of Sharma and coworkers, are absolutely capable of computing 

Concretely, we suggest the following procedure of handling these complex cases:
\begin{enumerate}
   \item Make an initial AVAS, with a target space including \emph{all} AOs suspected to be possibly relevant.
   \item Compute an approximate CASCI wave function $\ket{\Psi_\mathrm{approx}}$ in this active space---by using, for example, SHCI\cite{sharma:shci} (with high threshold $\epsilon$) or QC-DMRG\cite{sharma:QcdmrgSpinAdapted} (with low bond dimension $M$).
   \item Compute the natural orbitals $\{\psi_i\}$ and their occupation numbers $\{n_i\}$ from the approximate CASCI wave function $\ket{\Psi_\mathrm{approx}}$.
   \item Do the \emph{actual} quantitative calculation with an active space $\{\psi_i;\,t\leq n_i\leq (2-t)\}$, where $t$ is some numeric threshold (e.g., $t=0.05$). That is, use an active space composed of the natural orbitals $\psi_i$ from step 3 for which the occupation numbers $n_i$ differ significantly from 0.0 (empty in all configurations of $\ket{\Psi_\mathrm{approx}}$) and 2.0 (doubly-occupied in all configurations of $\ket{\Psi_\mathrm{approx}}$).
\end{enumerate}
While not investigated here, we expect this combination of methods to be capable of identifying suitable active spaces in many application scenarios.

\subsection{Quantitative treatment of homolytic bond dissociation processes}
The obtained data in Sec.~\ref{sec:FentonReaction} shows that the AVAS contains the necessary orbitals for describing the electron configurations relevant in a homolytic bond dissociation process, along the entire reaction path.
However, we note that for actual quantitative calculations, particularly in \emph{simple} concerted reaction processes  without hidden intermediates,\cite{kraka:UrvaReview,joo:MechanismOfBarrierlessReaction, kraka:StunningExampleOfComplexMechanism} modified procedures may be beneficial.
For example, if the chemical transformation of the reaction is sufficiently characterized by a single transition state (as in Sec.~\ref{sec:FentonReaction}), a more economical way of treating the reaction might be to compute an AVAS active space only at the transition state, use it to initialize a CASSCF calculation there, and then propagate the initial orbitals from geometry to geometry.
The main benefit of this latter approach would be that in the AVAS procedure, not only the rotated occupied and virtual orbitals with \emph{very low} overlap with the target AO space could be eliminated from the active space, but also the orbitals with \emph{very high} overlap (say, $\ge98\%$) could be eliminated, as such values indicate that they \emph{stay} doubly occupied/unoccupied during the reaction.
The active space size could therefore be substantially reduced in such cases, therefore allowing the application of more powerful electron correlation methods.
However, a detailed study of this is beyond the scope of this work.

We also wish to explicitly note that there are \emph{many} reactions in which the chemical transformations are \emph{not} characterized by only the transition state (e.g., see Ref.~\onlinecite{knizia:CurlyArrows} Fig.~3, or Refs.~\onlinecite{kraka:UrvaReview,joo:MechanismOfBarrierlessReaction,kraka:StunningExampleOfComplexMechanism}), and in these cases the ability to construct a consistent active space along the entire reaction path, such as afforded by the AVAS procedure, may be valuable.

}

% For example, a double-shell active space could be created by taking 
% ANO-RCC, ANO-VT-XZ or another non-minimal ANO basis set.
% For example, the ``double-shell' effect can be easily included in the AVAS method by including both 3$d$ and 4$d$ AOs in the ``target'' space of the active space construction
% using ANO-RCC basis set.}
% 
% If one wants to include additional AOs, which are not usually considered as valence AOs, but are believed 
% to be important for correlation, one should use ANO-RCC, ANO-VT-XZ or another non-minimal ANO basis set.
% For example, the ``double-shell' effect can be easily included in the AVAS method by including both 3$d$ and 4$d$ AOs in the ``target'' space of the active space construction
% using ANO-RCC basis set.}

% because this makes it pso
% 
% ``double-shell' effect can be easily included in the AVAS method by including both 3$d$ and 4$d$ AOs in the ``target'' space of the active space construction
% using ANO-RCC basis set.
% 
% 
% If one wants to include additional AOs, which are not usually considered as valence AOs, but are believed 
% to be important for correlation, one should use ANO-RCC, ANO-VT-XZ or another non-minimal ANO basis set.
% For example, the ``double-shell' effect can be easily included in the AVAS method by including both 3$d$ and 4$d$ AOs in the ``target'' space of the active space construction
% using ANO-RCC basis set.}

\section{Conclusions}\label{sec:Conclusions}

In this work, we investigated how to systematically and automatically construct
molecular active spaces solely from a single determinant wavefunction
together with a list of atomic valence orbitals. The atomic valence active space (AVAS) procedure is based on a straightforward
linear algebraic rotation of the occupied and unoccupied molecular orbital spaces which maximizes their given atomic valence character.
The method automatically detects the valence bonding partners of a given atomic valence orbital, and, by using 
a single small threshold, can also detect non-bonding orbitals without constructing
spurious partners (either occupied or unoccupied).

To test our scheme, we tested both the quality of our orbitals as initial guesses for CASSCF optimization,
as well as the accuracy and stability of the valence excitation energies calculated within our spaces. We find
high overlap of our orbitals with fully optimized CASSCF orbitals, demonstrating their high quality.
We can also obtain good CASSCF excitation energies in cases where the excitations are dominated by valence correlation.
In molecules where the excitations are not of this character, we find that the
addition of dynamic correlation (through the $N$-electron valence perturbation theory)
yields quantitative agreement with experiment.

No doubt it will still be necessary to experiment with active spaces in the modeling of large and very complex molecules.
Our study  provides two reasons to believe that the difficulty of
performing reliable multireference calculations in complex problems can be reduced  using the AVAS technique.
First,  the simple procedure makes it trivial to obtain not only the minimal active spaces, but 
also extended active spaces, for example, including additional ligand orbitals.
This makes it simpler to systematically explore different active spaces,  eliminating user error and subjectivity in their definition,
and allowing for convergence of properties with respect to the active space size.
Second, systematically varying active space size, while including a dynamic correlation treatment (such as NEVPT2) 
provides a straightforward way to  assess whether our active space is converged, as computed observables should become
insensitive to the active space size.
For these reasons, we believe that the AVAS construction provides a simple route to painless multireference calculations by non-experts,
particularly in complex systems involving transition metals.

\section*{Supporting Information}
Supporting information: Additional computational details; geometric structures of the studied transition metal complexes; additional figures of studied active orbital spaces; numerical and structural data of the studied reaction path of the Fenton reaction.
This material is available free of charge via the Internet at http://pubs.acs.org.

\section*{Acknowledgments}
We acknowledge the US National Science Foundation for funding this research primarily through the award NSF:CHE-1665333. Additional support for software development and to support QS was provided through NSF:CHE-1657286. We 
acknowledge additional support for GKC from the Simons Foundation through a Simons Investigatorship.

\bibliography{refs}
% ^- this refers to your bibtex file (refs.bib in this case).

\end{document}